\documentclass[aps,physrev,10pt,floatfix,longbibliography,nofootinbib,reprint]{revtex4-2}
\usepackage{algorithm}
\usepackage{algpseudocode}
\usepackage{amsmath}
\usepackage{amssymb}
\usepackage[english]{babel}
\usepackage{bm}
\usepackage{cancel}
\usepackage{graphicx}
\usepackage{hyperref}
\usepackage{xcolor}

\newcommand{\bmpi}{{\bm{\pi}}}

\newcommand{\e}{\mathrm{e}}
\newcommand{\s}{{\mathbf{s}}}
\newcommand{\Var}[1]{\mathop{\mathrm{Var}}\left[ #1 \right]}

\newcommand{\Alg}[1]{Alg.~\ref{alg:#1}}
\newcommand{\Eq}[1]{Eq.~(\ref{eq:#1})}
\newcommand{\Fig}[1]{Fig.~\ref{fig:#1}}

\begin{document}

\title{Unbiased Monte Carlo cluster updates with autoregressive neural networks}
\author{Dian Wu} \email{dian.wu@epfl.ch}
\author{Riccardo Rossi} \email{riccardo.rossi@epfl.ch}
\author{Giuseppe Carleo} \email{giuseppe.carleo@epfl.ch}
\affiliation{Institute of Physics, \'Ecole Polytechnique F\'ed\'erale de Lausanne (EPFL), CH-1015 Lausanne, Switzerland}
\date{\today}

\begin{abstract}
Efficient sampling of complex high-dimensional probability distributions is a central task in computational science. Machine learning methods like autoregressive neural networks, used with Markov chain Monte Carlo sampling, provide good approximations to such distributions, but suffer from either intrinsic bias or high variance. In this Letter, we propose a way to make this approximation unbiased and with low variance. Our method uses physical symmetries and variable-size cluster updates which utilize the structure of autoregressive factorization. We test our method for first- and second-order phase transitions of classical spin systems, showing its viability for critical systems and in the presence of metastable states.
\end{abstract}

\maketitle

\section{Introduction}

Markov chain Monte Carlo~\cite{mcmc} (MCMC) is an unbiased numerical method that allows sampling from unnormalized probability distributions, a central task in many areas of computational science. MCMC is commonly used, for example, in molecular dynamics~\cite{binder1995monte}, as well as statistical and quantum physics~\cite{binder2010monte, krauth2006statistical, werner_book, sorella_book}. In addition to fundamental applications, MCMC serves as a physics-inspired approach to solve a variety of computational problems, including combinatorial optimization~\cite{kirkpatrick1983optimization, rubinstein2013cross} and computer graphics~\cite{cook1986stochastic}. While MCMC is a generically applicable technique, its implementation can be plagued by long mixing or autocorrelation time~\cite{muller1973dynamic}. Various techniques have been proposed to increase the efficiency of MCMC~\cite{liu2008monte}, for example, cluster updates~\cite{wang1990cluster, wolff1989collective}, parallel tempering~\cite{swendsen1986replica}, the worm algorithm~\cite{prokofev1998worm}, and event-chain Monte Carlo~\cite{ecmc_krauth}. However, these faster MCMC algorithms rely on details of the physical system considered, and they cannot be applied generically.

Machine learning (ML) methods, given their intrinsic flexibility in addressing problems in computational physics~\cite{carleo2019machine}, are being intensively investigated as a way to improve MCMC. Applications in this direction include, for example, self-learning Monte Carlo methods~\cite{levy2017generalizing, song2017anicemc, medvidovic2020generative, liu2017self, huang2017accelerated, shen2018self}, enhanced sampling driven by neural networks~\cite{bonati2019neural, noe2019boltzmann}, and neural importance sampling~\cite{muller2019neural}. Strongly rooted in the principles of statistical physics, variational sampling techniques are among the most promising ML-driven approaches. Generative neural samplers (GNS)~\cite{wu2019solving, albergo2019flow, li2018neural} are a chief example of ML-driven variational methods. These approaches build on the idea of constructing approximate representations of the original probability distribution at hand. The resulting variational approximations can efficiently perform sampling by construction, thus completely bypassing MCMC. A particularly interesting aspect of this approach is its systematic improvability when using the free energy bound minimization as the guiding principle to gauge the approximation accuracy. The main drawback of the variational approach, however, is that the estimators of expectation values are intrinsically biased by the representation error of the approximated distribution. As unbiased estimators are of central importance in many fundamental applications in physics, recent research has started addressing the key problem of removing the bias induced by ML variational representations, for example, through importance sampling, and incorporating again MCMC strategies~\cite{muller2019neural, nicoli2020asymptotically, mcnaughton2020boosting}.

In this Letter, we propose a way to combine variational techniques with MCMC by using autoregressive neural networks~\cite{uria2016neural, pmlr-v15-larochelle11a} to propose cluster updates. We first show that existing unbiased sampling schemes using global updates proposed by GNS can be plagued by the ergodicity issue due to the generic presence of ``exponentially suppressed configurations'', which have a limited effect on the variational free energy but a rather strong effect on the autocorrelation time. Our workaround to this problem consists of two ingredients. On one hand, we consider physical symmetry operations that leave the Hamiltonian invariant. When applied to the MCMC states, these symmetry operations significantly reduce the exponential suppression of configurations belonging to the same equivalence class. On the other hand, we take advantage of the structure of autoregressive factorization to propose MCMC updates with clusters of spins. The cluster update scheme is automatically learned for any Hamiltonian and is therefore particularly helpful for Hamiltonians with no known cluster update scheme. We benchmark our technique on the two-dimensional Ising model, showing that our solution eliminates the ergodicity issue of the global update approach in the critical region. We then study an Ising-like frustrated plaquette model for which traditional cluster algorithms are not applicable, and we find a first-order transition from a paramagnetic state to a ``ferrimagnetic'' state that breaks the $\mathbb{Z}_2 \times \mathbb{Z}_2 \times \mathbb{Z}_2$ symmetry of the Hamiltonian. We show that the method greatly alleviates the metastability issue, as it can rapidly thermalize by cluster updates.

\subsection{Bias in neural sampling}

In the following we consider a system of $V$ classical Ising spins $\s := (s_1, \dots, s_V)$, $s_i \in \{-1, 1\}$, at inverse temperature $\beta$. We use a GNS $q_\theta$ with parameters $\theta$ that variationally approximates the Boltzmann probability distribution $p(\s) \propto \tilde{p}(\s) := \e^{-\beta E(\s)}$ by minimizing, in the language of statistical physics, a free energy bound~\cite{wu2019solving}, which is equivalent to minimizing the Kullback--Leibler (KL) divergence~\cite{kullback1951information}:
\begin{equation}
D_\text{KL}(q_\theta \,\|\, p) := \sum_\s q_\theta(\s) \ln \frac{q_\theta(\s)}{p(\s)}.
\label{eq:kl}
\end{equation}
To construct an expressive $q_\theta$, we use an autoregressive neural network to decompose it into a product of conditional probabilities $q_\theta(\s) =: \prod_{i = 1}^V q_{\theta; i}(s_i \mid \s_{< i})$, where $\s_{< i} := (s_1, \ldots, s_{i - 1})$. This specific choice for the model allows us to efficiently sample from the distribution $q_\theta(\s)$ by sampling from the conditional probabilities $\{q_{\theta; i}\}$ sequentially~\cite{wu2019solving}.

The variational autoregressive approach is systematically improvable and allows exact sampling. However, the fact that the two distributions are only approximately equal, $q_\theta(s) \approx \tilde{p}(\s)$, also implies that the samples $\{\s^{(1)}, \ldots, \s^{(N)}\}$ drawn from the network carry an intrinsic bias. When these samples are used to compute the expectation value of an observable, the resulting estimator $\bar{O} = \frac{1}{N} \sum_{i = 1}^N O(\s^{(i)})$ is biased, and most importantly, it is not possible in general to reliably estimate the direction and the magnitude of such bias.

\subsection{Neural importance sampling and global updates}

Refs.~\cite{nicoli2020asymptotically, mcnaughton2020boosting} have proposed two closely related solutions to the bias problem. The first method, which we denote neural importance sampling (NIS) in the following, consists of using the modified unbiased estimator $\bar{O} = \sum_{i = 1}^N w(\s^{(i)}) O(\s^{(i)})$, where $w(\s^{(i)}) := \frac{\tilde{w}(\s^{(i)})}{\sum_{j = 1}^N \tilde{w}(\s^{(j)})}$ and $\tilde{w}(\s^{(i)}) := \frac{\tilde{p}(\s^{(i)})}{q_\theta(\s^{(i)})}$ are the normalized and the unnormalized weights respectively. The second proposed solution, which we denote neural global updates (NGU) hereafter, consists of using the GNS as a MCMC proposer: if $\s$ is the Markov chain state, a proposed state $\s'$ is drawn from the GNS and accepted with the Metropolis probability:
\begin{equation}
P_{\text{acc}}(\s \to \s') := \min \left( 1, \frac{\tilde{w}(\s')}{\tilde{w}(\s)} \right).
\label{eq:P_a}
\end{equation}

\section{Methods}

\subsection{Exponentially suppressed configurations}

We point out an elementary property of the KL divergence, \Eq{kl}: the cost of allowing a single bad approximation $q_\theta(\s)$ scales only logarithmically with the ratio $q_\theta(\s) / p(\s)$. Therefore, it is reasonable to expect that, even when the free energy is well approximated after the variational training, $q_\theta(\s)$ is still exponentially smaller than $p(\s)$ for a small portion of configurations. We call them exponentially suppressed configurations (ESC)~\footnote{In the context of variational inference, it was empirically found~\cite{goodfellow2016deep} that $q$ tends to cover fewer modes than $p$ in the probability landscape. However, the problem of exponentially suppressed configurations is more general as it is present even when all the modes are represented.}. We denote $p_\text{ESC}$ to be the probability that a configuration is an ESC~\footnote{A distribution of $\tilde{w}$ is shown in Fig.~S3 in the Supplemental Material~\cite{sm}, which can be used to estimate $p_\text{ESC}$.}. A well-trained network has $p_\text{ESC} \ll 1$, and they have a limited effect on the variational free energy but a rather strong effect on the autocorrelation time.

Let us consider a Markov chain evolution using NGU, and suppose that the Markov chain state $\s$ is an ESC. The ratio $\tilde{w}(\s') / \tilde{w}(\s)$ in \Eq{P_a} will be exponentially small for almost any other configuration $\s'$; therefore, the Markov chain will be essentially stuck in $\s$ for a long time before accepting any new proposal, and the autocorrelation time of the whole chain will be impractically large. A similar argument applies when considering the variance of the NIS method.

\subsection{Symmetry-enforcing updates}

To solve the generic ergodicity problem of neural global update methods, we start by proposing an enhanced MCMC method to enforce the symmetries. At each Monte Carlo step, we apply a random element of the symmetry group $G$, composed of a translation and reflections, to the current configuration~\footnote{Formally, this is equivalent to multiplying the Markov transition matrix by another matrix that leaves the equilibrium distribution invariant, as discussed in the Supplemental Material~\cite{sm}.}. There is no need to reject this action because the energy is invariant under the action. In the following, we refer to this method as neural global updates with symmetries (NGUS).

Assume that the current configuration $\s$ is an ESC, and we use a random symmetry operation to change $\s$ to another configuration $\s^*$ in the equivalence class $\mathcal{C}$. The probability that all configurations in $\mathcal{C}$ are ESC is on the order of $p_\text{ESC}^{\#\mathcal{C}}$, so it is extremely unlikely to get stuck within the equivalence class. The occurrence of ESC does not depend on the physical symmetries but rather the structure of the network.

\subsection{Neural cluster updates}

With autoregressive neural networks, it is particularly natural to consider cluster updates where only a subset of the lattice is changed. Indeed, for any given $k$, it is possible to propose an update $\s \to \s'$ by setting $\s'_{\le V - k} := \s_{\le V - k}$ and only sample $\s'_{> V - k}$. The weight ratio in \Eq{P_a} becomes $\frac{\tilde{w}(\s')}{\tilde{w}(\s)} = \frac{\tilde{p}(\s')}{\tilde{p}(\s)} \prod_{i = V - k + 1}^V \frac{q_{\theta; i}(s_i \mid \s_{< i})}{q_{\theta; i}(s'_i \mid \s'_{< i})}$, which is not too far from $1$ when $k$ is small. In this way, the new configuration is closer to the old one and is easier to be accepted, so we expect lower autocorrelation time than global update methods. In the following, we refer to this method as neural cluster updates (NCU).

\begin{algorithm}[H]
\caption{A step of NCUS.}
\label{alg:ncus}
\begin{algorithmic}[1]
\State Input the current configuration $\s$
\State Sample an integer $k \in \{1, \ldots, V\}$ from $P_\text{cluster}$
\State Sample the last $k$ spins and propose the configuration $\s'$
\State Accept $\s \gets \s'$ with probability $P_\text{acc}(\s \to \s')$
\State Translate $\s$ by a random displacement
\State Reflect $\s$ along the $x$ axis, the $y$ axis and the diagonal, each with $50\%$ probability
\State Reflect $\s$ along the $z$ axis (flip all spins) with $50\%$ probability
\State Output $\s$ as a sample in the Markov chain
\end{algorithmic}
\end{algorithm}

\begin{figure}[htb]
\includegraphics[width=\linewidth]{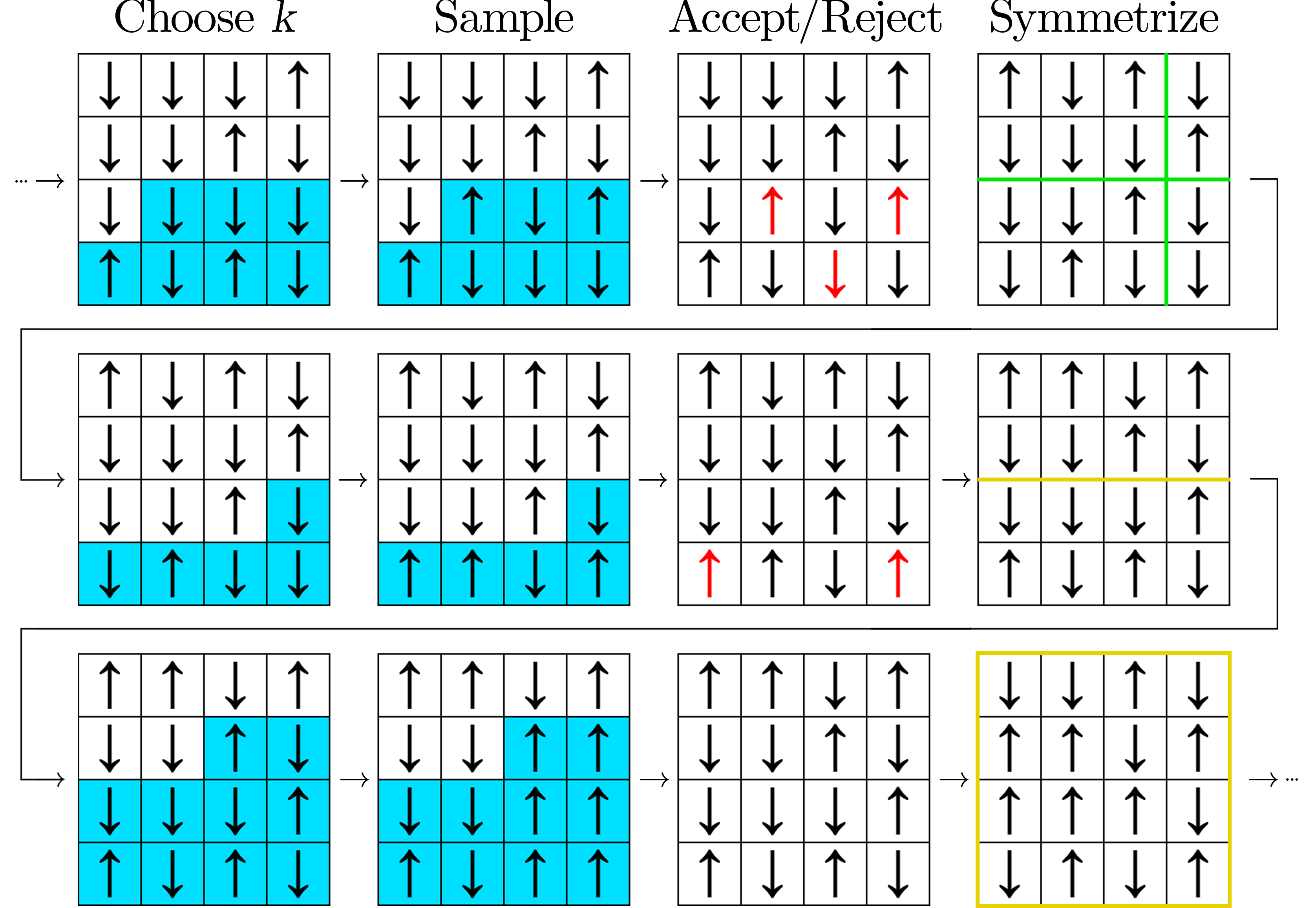}
\caption{Example of three steps of NCUS applied to a $4 \times 4$ spin model. The columns correspond to different lines in \Alg{ncus}. The last $k$ spins that can be flipped are highlighted in {\color[HTML]{0099DD} blue}. If a proposal is accepted, the spins actually flipped are shown in {\color[HTML]{DD0000} red}. For translations, the original borders of the lattice are shown in {\color[HTML]{00AA00} green}. For reflections, the plane of reflection is shown in {\color[HTML]{DDAA00} yellow}, and yellow borders around the lattice indicate a reflection along the $z$ axis (across the $x y$ plane).}
\label{fig:ncus_proc}
\end{figure}

As symmetry-enforcing and cluster updates are compatible with each other, we use the two at the same time and we call the resulting method neural cluster updates with symmetries (NCUS), which still falls into the category of MCMC. As described in \Alg{ncus}, we randomly choose a cluster size $k$ and consider the last $k$ spins $\s_{> V - k}$ in the autoregressive order~\footnote{The autoregressive order is a one-dimensional labeling $s_k$ of the spins in the lattice, mapped from the two-dimensional labeling $s_{i, j}$, where $k = (i - 1) \times L + j$.} to be inside the cluster. Then we sample those spins from the approximate distribution $q_\theta$ of the physical system learned into the network, which is conditioned on the spin configuration $\s_{\le V - k}$ outside the cluster. After that, we accept the new configuration $\s'$ with the probability $P_\text{acc}(\s \to \s')$, then randomly apply the symmetry operations. See \Fig{ncus_proc} for a schematic illustration.

Although the cluster size $k$ can come from an arbitrary distribution $P_\text{cluster}(k)$, numerical experiments have shown that the uniform distribution $P_\text{cluster}(k) \equiv 1/V$ already works better than many other cases we have explored~\footnote{A comparison of different choices of $P_\text{cluster}$ can be found in Figs.~S1 and S2 in the Supplemental Material~\cite{sm}.}.

\begin{figure*}[htb]
\includegraphics[width=\linewidth]{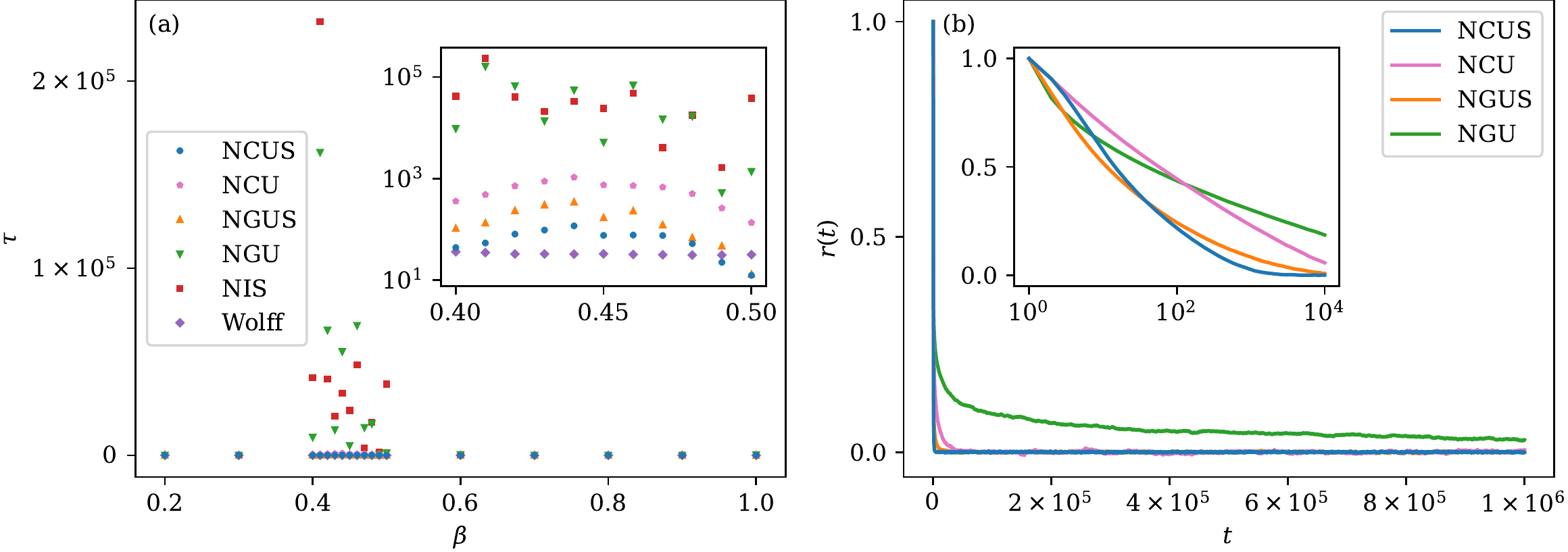}
\caption{(a) Integrated autocorrelation time $\tau$ as a function of temperature on the $16 \times 16$ Ising model. For neural importance sampling (NIS), we use the increased variance from the reweighting procedure as the effective autocorrelation time. The inset focuses on their behaviors near the critical point and uses the logarithmic scale on the $y$ axis. (b) Autocorrelation functions $r(t)$ on the $16 \times 16$ Ising model at $\beta = 0.44$. The inset uses the logarithmic scale on the $x$ axis to focus on their behaviors at small $t$.}
\label{fig:ising_autocorr}
\end{figure*}

\section{Numerical experiments}

\subsection{Ising model}

We start to demonstrate the effectiveness of NCUS on the conventional two-dimensional Ising model:
\begin{equation}
E(\s) := \sum_{i, j = 1}^L s_{i, j} (s_{i + 1, j} + s_{i, j + 1}),
\end{equation}
with periodic boundary conditions $s_{L + 1, j} = s_{1, j}$, $s_{i, L + 1} = s_{i, 1}$. The model can be solved exactly and has a critical point at $\beta = \ln(1 + \sqrt{2}) / 2 \approx 0.44$~\cite{onsager1944crystal}.

Our network architecture is based on PixelCNN~\cite{van2016pixel}, combined with dilated convolutions~\cite{yu2016multi} to reduce the total number of parameters. Overall, our networks are lightweight and have $3$ convolutional layers and approximately $4 \times 10^3$ parameters. Thanks to the MCMC bias removal, we do not need the network to approximate the true distribution to extremely high precision, which in any case will be increasingly difficult for larger lattices. As we use the same network for all the experiments, we can compare the performances of the various unbiased sampling methods. After training the network, we generate $10^3$ Markov chains in parallel, each containing $10^5$ samples~\footnote{Details of the network structure, training, and sampling are described in the Supplemental Material~\cite{sm}.}.

When comparing the efficiencies of different MCMC algorithms, the main metric is the integrated autocorrelation time $\tau$~\footnote{For completeness, we provide the definitions of the autocorrelation function and the integrated autocorrelation time in the Supplemental Material~\cite{sm}.}, which determines the variance of the estimator when the variance of the observable and the sample size are given. Here, $\tau$ is an intrinsic property of the algorithm and the physical system, without dependence on the sample size, if we have enough samples to obtain a converged estimation of it. For NIS, the autocorrelation time is equal to one by definition; however, there is an increased variance arising from the reweighting procedure, which we consider an effective autocorrelation time for the sake of comparison with the other techniques.

From \Fig{ising_autocorr}~(a), we see that both NGU and NIS have pathologically high autocorrelation times in the critical region. An inspection of their autocorrelation function [see \Fig{ising_autocorr}~(b)] shows that the Markov chain of NGU is essentially nonergodic in the available simulation time. By contrast, our proposed method NCUS has no issue in the critical region. A closer inspection of the inset of \Fig{ising_autocorr}~(a) shows that the autocorrelation time of NCUS still increases in the critical region, and the sampling efficiency is improved typically by $2$ orders of magnitude compared with the global update methods. The performance of NCUS is also comparable with the celebrated Wolff cluster update method~\cite{wolff1989collective}, which is specifically tailored for the Ising model. Both NCUS and NGUS perform well in the critical region, and NCUS is to be preferred, as the cluster update allows us to achieve a lower autocorrelation time and, more importantly, a better asymptotic behavior of the autocorrelation function.

\begin{figure*}[htb]
\raisebox{10pt}{\includegraphics[width=0.49\linewidth]{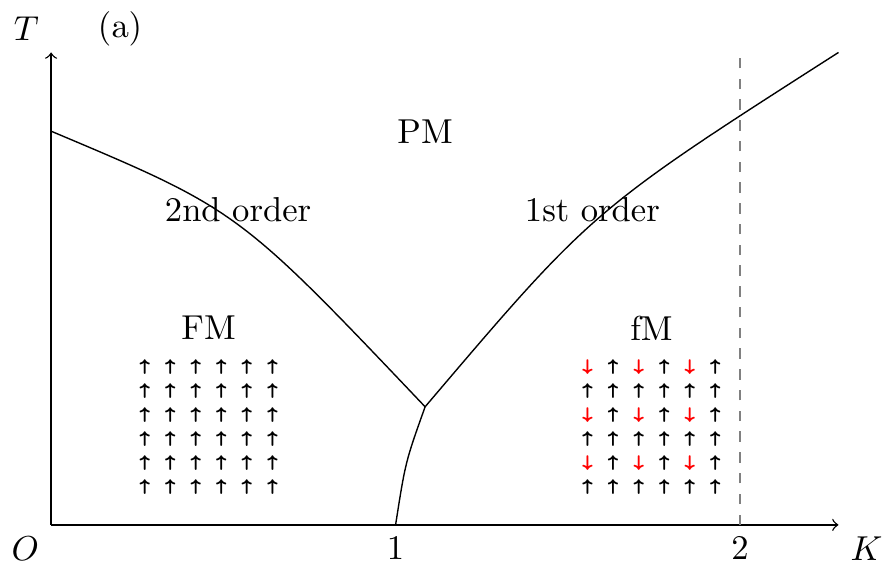}}
\includegraphics[width=0.49\linewidth]{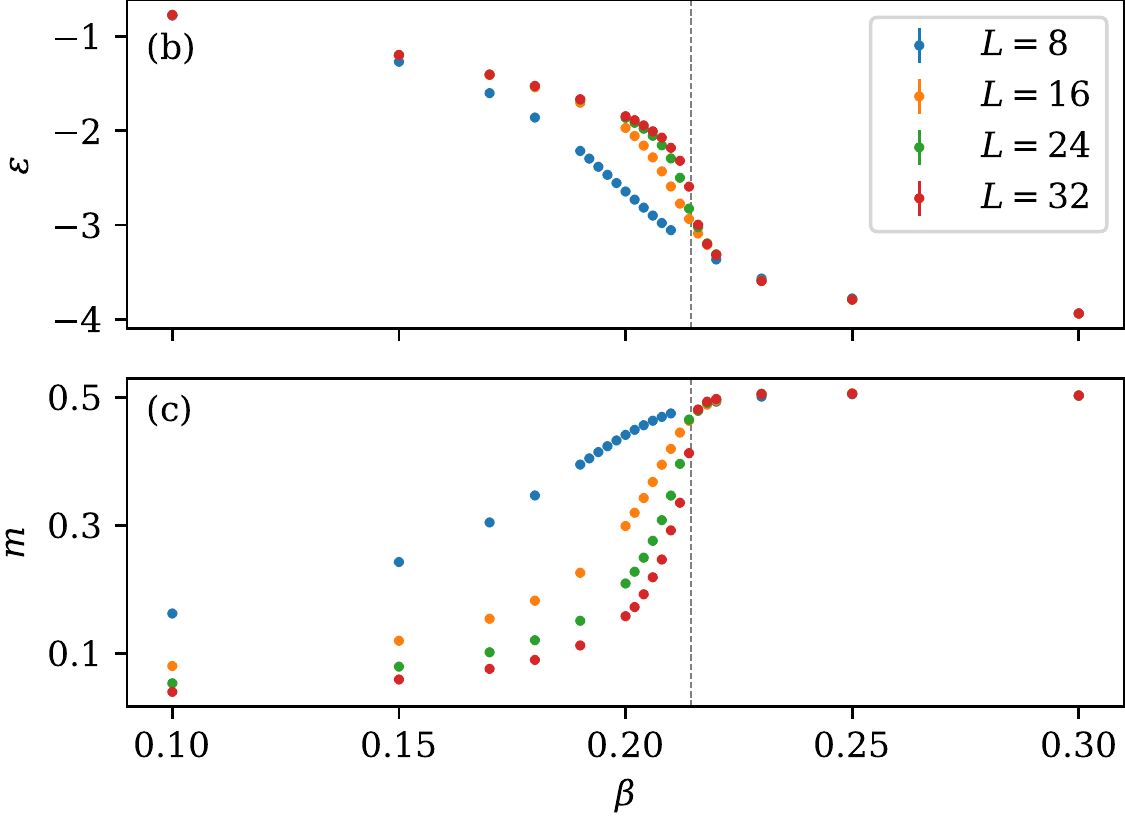}
\caption{(a) Sketch of the expected phase diagram of the frustrated plaquette model (FPM) for $J_1 = J_3 = -1$ with example ground states in the ferromagnetic (FM) and the ferrimagnetic (fM) phases. The dashed vertical line represents the temperature cut we numerically study. (b) Energy per site $\varepsilon$ and (c) spontaneous magnetization per site $m$ of the FPM for $J_1 = J_3 = -1$, $K = 2$ as functions of temperature with lattice sizes up to $L = 32$, obtained by neural cluster updates with symmetries (NCUS). The dashed vertical line indicates the estimated transition point $\beta_c$.}
\label{fig:fpm}
\end{figure*}

\begin{figure*}[htb]
\includegraphics[width=\linewidth]{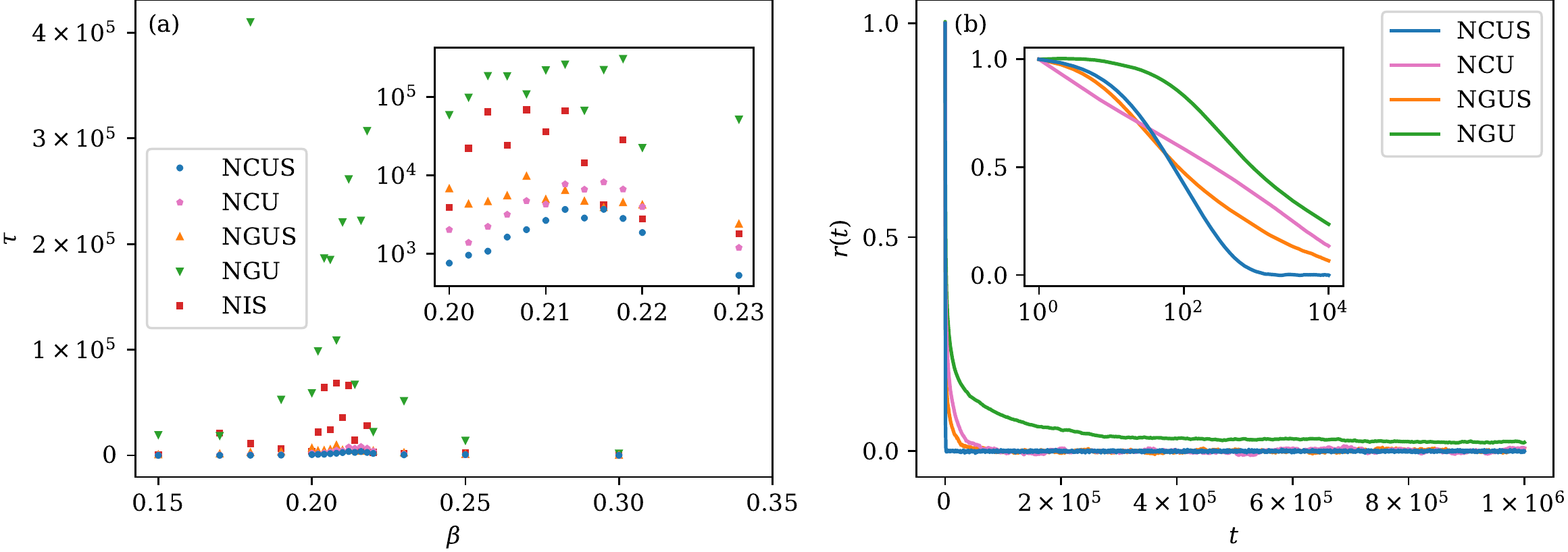}
\caption{(a) Integrated autocorrelation time $\tau$ as a function of temperature on the $32 \times 32$ frustrated plaquette model (FPM). For neural importance sampling (NIS), we use the increased variance from the reweighting procedure as the effective autocorrelation time. The inset focuses on their behaviors near the critical point and uses the logarithmic scale on the $y$ axis. (b) Autocorrelation functions $r(t)$ on the $32 \times 32$ FPM at $\beta = 0.2$. The inset uses the logarithmic scale on the $x$ axis to focus on their behaviors at small $t$.}
\label{fig:fpm_autocorr}
\end{figure*}

\subsection{Frustrated plaquette model}

We now study another model that presents a richer physics than the Ising model and for which, to our knowledge, no traditional cluster update method is applicable. We consider a classical spin-$1/2$ system with nearest-neighbor $J_1$, next-next-nearest-neighbor $J_3$, and plaquette $K$ interactions:
\begin{align}
E(\s) := &\phantom{{}+{}} J_1 \sum_{i, j = 1}^L s_{i, j} (s_{i + 1, j} + s_{i, j + 1}) \nonumber \\
&+ J_3 \sum_{i, j = 1}^L s_{i, j} (s_{i + 2, j} + s_{i, j + 2}) \nonumber \\
&+ K \sum_{i, j = 1}^L s_{i, j} \, s_{i + 1, j} \, s_{i, j + 1} \, s_{i + 1, j + 1},
\end{align}
with periodic boundary conditions, which we denote as the frustrated plaquette model (FPM). In this Letter, we set $J_1 = J_3 = -1$ and $K > 0$.

We sketch the expected phase diagram in \Fig{fpm}~(a). The ground state of the FPM depends on the competition of $J_1$ and $K$. For small $K$, we expect a transition as a function of temperature between a paramagnetic (PM) and a ferromagnetic phase (FM), which is analogous to what is found in the conventional Ising model. For $K > 1$, the ground state is a repetition of a $2 \times 2$ unit cell containing one spin pointing in the opposite direction of the other three spins, as shown in \Fig{fpm}~(a). The ground state breaks the $\mathbb{Z}_2$ spin-inversion symmetry and the $\mathbb{Z}_2 \times \mathbb{Z}_2$ translation symmetry of one lattice spacing in $x$ and $y$ directions. As this phase has an average magnetization per site of $1/2$, we refer to this phase as ferrimagnetic (fM). At finite temperature, there must be a phase transition between the PM phase and the fM one, the nature of which we investigate in this Letter.

We present the numerical results for the energy per site $\varepsilon$ and the spontaneous magnetization per site $m := \left\langle \left\lvert \frac{1}{V} \sum_i s_i \right\rvert \right\rangle$ in \Fig{fpm}~(b,~c) respectively, as functions of temperature with $K = 2$ and lattice sizes up to $L = 32$. Our results in \Fig{fpm}~(b) strongly suggest that, in the thermodynamic limit $L \to \infty$, the energy is discontinuous at the critical point $\beta_c = 0.2145 \pm 0.0012$ with a latent heat $Q = 1.36 \pm 0.20$, estimated using the standard finite-size scaling procedure of Ref.~\cite{vollmayr1993finite}. Another indication of the first-order nature of the phase transition comes from the spontaneous magnetization shown in \Fig{fpm}~(c), which when extrapolated to the thermodynamic limit shows a discontinuity of the spontaneous magnetization from 0 to a value close to $1/2$, as expected for the fM phase~\footnote{A naive Ginzburg--Landau approach for a three-component $\mathbb{Z}_2 \times \mathbb{Z}_2 \times \mathbb{Z}_2$-symmetric order parameter predicts a second-order transition when truncated at the quartic level. A first-order phase transition is also found in the $q = 8$ Potts model~\cite{arisue2001first, gorbenko2018walking, gorbenko2018walking2}, but the broken symmetry group there is $\mathbb{Z}_8$.}.

\begin{figure}[htb]
\centering
\includegraphics[width=\linewidth]{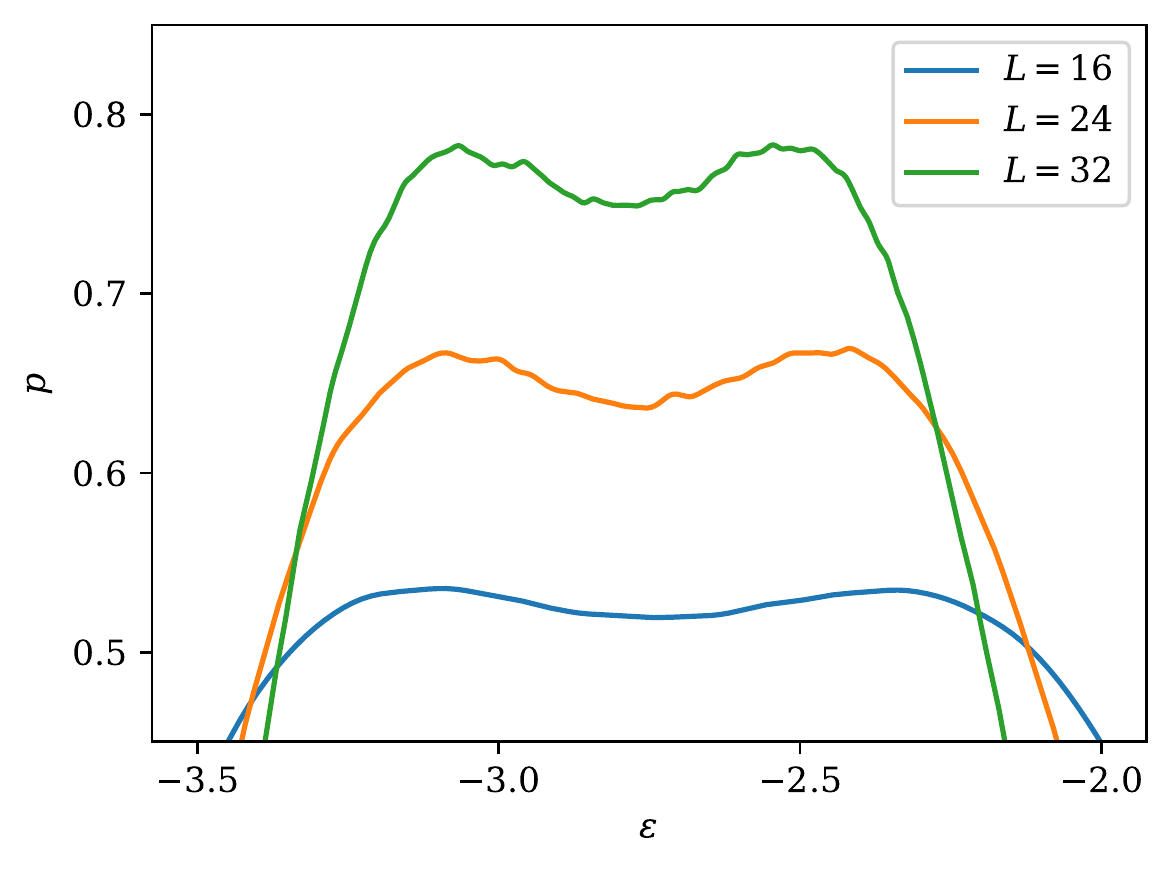}
\caption{Probability distribution of the energy per site $p(\varepsilon)$ for the FPM with different lattice sizes $L$ at their respective phase transition temperatures, obtained by NCUS.}
\label{fig:fpm_hist_p}
\end{figure}

The comparison of autocorrelation times from different sampling methods is presented in \Fig{fpm_autocorr}, which provides numerical evidence that NCUS greatly alleviates the metastability issue expected near first-order phase transitions~\cite{gheissari2016mixing, krzakala2009hiding, krzakala2012statistical, banks2017lovasz}. Theoretically, a first-order phase transition occurs when the distribution of energy $p(E) \propto N(E) \, \e^{-\beta E}$ has two peaks with the same size, as shown in Fig.~\ref{fig:fpm_hist_p}, where $N(E)$ is the number of configurations with energy $E$. A GNS-based sampling method has equal probabilities to generate a sample from the two peaks, and the probability to accept that proposal will be close to $1$, if the network is ideally trained and there is no problem of ESC. Meanwhile, for traditional local-update MCMC methods, they can only move small horizontal steps in Fig.~\ref{fig:fpm_hist_p}, so it takes more steps ($\sim L^2$) and exponentially lower probability ($\sim \e^{-\beta \delta E L^2}$) for them to walk from the low-energy peak to the high-energy one, where $\delta E$ is the typical energy difference in a local update, which does not scale with $L$. In other words, the exponentially large number of configurations in the high-energy peak will not make it easier for local-update MCMC methods to sample from that peak because it is exponentially hard for the walker to walk between those configurations in locally connected paths. NCUS reaches a balance between the two extremes, which solves the problem of ESC and keeps the autocorrelation time practically low, even if the network is lightweight and cannot ideally approximate the true distribution.

Another potential issue in first-order phase transitions is the strong divergence of the specific heat, resulting in high variance of the energy. Despite this, NCUS still helps us estimate the energy with high accuracy and the error bars in \Fig{fpm}~(b,~c) are too small to be visible.

\section{Conclusions}

In this Letter, we have shown a strategy to systematically remove the bias of variational autoregressive neural network methods and, at the same time, keep the variance of observables under control. Our approach exploits the autoregressive structure of the models to generate cluster Monte Carlo updates. After having shown that global updates proposed from networks trained with the KL divergence are generically expected to fail because of a small number of exponentially suppressed configurations (ESC), we have provided a workaround that takes advantage of enforcing the symmetries of the physical system and from using the chainlike graphical structure of the autoregressive model, namely NCUS, to help the Markov chain rapidly escape from ESC. We have benchmarked our technique for the two-dimensional Ising model, showing its efficacy in the critical region, where a straightforward implementation of neural global updates fails. We have further shown the potential of our method for systems for which no traditional cluster updates are known by considering a frustrated plaquette Ising model, where we were able to determine the first-order nature of a paramagnetic-ferrimagnetic phase transition breaking a $\mathbb{Z}_2 \times \mathbb{Z}_2 \times \mathbb{Z}_2$ symmetry, remarking that the automatic cluster updates we used allowed us to alleviate the metastability issue.

While we have been mainly concerned with the metric of autocorrelation time, we recognize that the wall-clock time is another important metric for practical computations. In this respect, when computing the energy of the system has a negligible computational cost, current neural network-based methods are not yet competitive with traditional MCMC methods. It can then be argued that the ideal application scenario for ML-based methods are those cases where evaluating the integrand is expensive, for example, in determinant quantum Monte Carlo~\cite{blankenbecler1981monte} and lattice field theory~\cite{hackett2021flow}. In future work, computational efficiency can be addressed on multiple fronts, for example, by introducing techniques such as hierarchy and sparsity of the neural network models, to reduce the computation time and scale up the lattice size by orders of magnitude. After that, we expect that the slow asymptotic growth of the autocorrelation time of GNS will eventually make them outperform traditional MCMC methods in terms of wall-clock time.

\begin{acknowledgments}
We acknowledge insightful comments and suggestions from Lei Wang and Pan Zhang. Support from the Swiss National Science Foundation is acknowledged under Grant No. 200021\_200336. The computing power is supported with Cloud TPUs from Google's TensorFlow Research Cloud (TFRC). Our code is available at: \url{https://github.com/wdphy16/neural-cluster-update}
\end{acknowledgments}

\clearpage
\widetext

\begin{center}
\textbf{\large Supplemental material}
\end{center}

\setcounter{section}{0}
\setcounter{equation}{0}
\setcounter{figure}{0}
\setcounter{table}{0}

\renewcommand{\thesection}{S\arabic{section}}
\renewcommand{\theequation}{S\arabic{equation}}
\renewcommand{\thefigure}{S\arabic{figure}}
\renewcommand{\thetable}{S\arabic{table}}

\section{Decomposition of transition matrix}

The transition matrix $M$ of a Markov chain is defined by
\begin{equation}
\bmpi_{t + 1} = M\,\bmpi_t,
\end{equation}
where $\bmpi_t$ is a state vector containing the probabilities of $2^V$ configurations at sampling time $t$. $M_{i j}$ is the probability to move from the configuration $j$ to $i$. $M$ has the \emph{left stochastic property}
\begin{equation}
\sum_i M_{i j} = 1, \quad \forall j.
\end{equation}

To improve the acceptance rate of the chain, we can write $M$ as a convex combination of two transition matrices
\begin{equation}
M = \lambda M^{(1)} + (1 - \lambda) M^{(2)},
\end{equation}
or as a product of two transition matrices
\begin{equation}
M = M^{(2)} M^{(1)},
\end{equation}
where $M^{(1)}$ and $M^{(2)}$ are easier to sample than $M$. Given $M^{(1)}$ and $M^{(2)}$ are transition matrices, $M$ must be a transition matrix and hold the left stochastic property.

\begin{figure*}[htb]
\includegraphics[width=\linewidth]{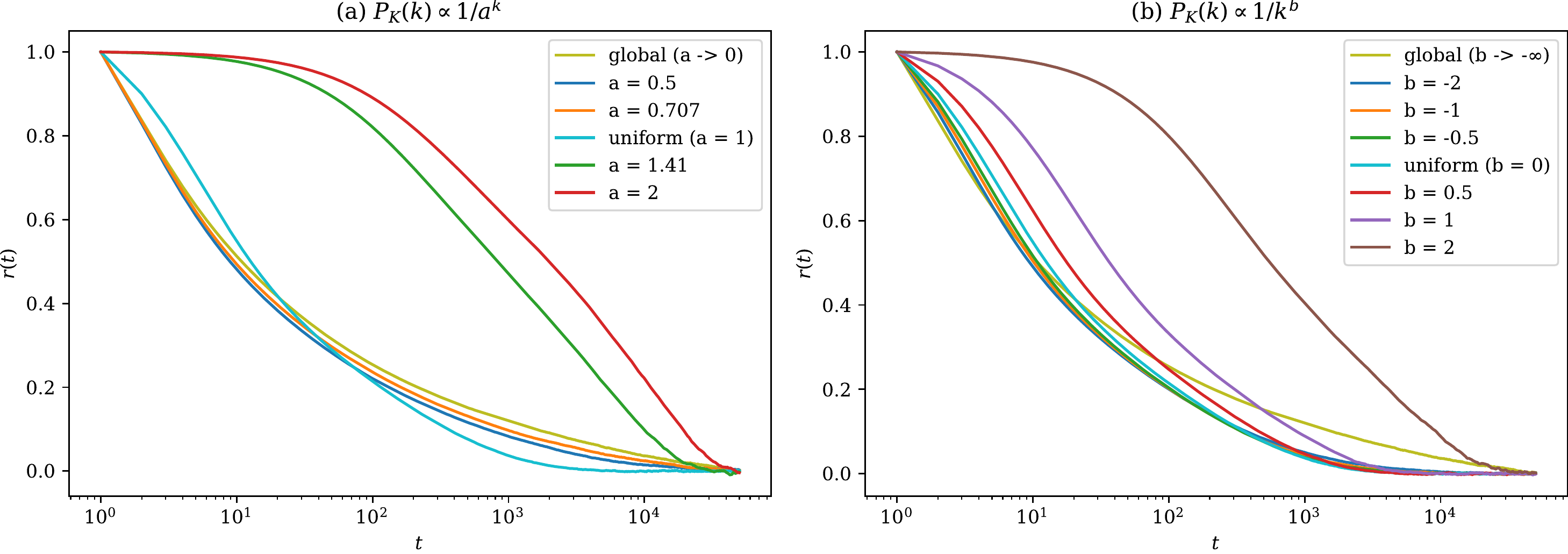}
\caption{Autocorrelation functions $r(t)$ by NCUS with various exponential and power distributions $P_\text{cluster}$, computed on the $16 \times 16$ Ising model. In general, the uniform distribution gives lower IAT than most other distributions we experimented.}
\label{fig:ising_autocorr_fun_dist_k}
\end{figure*}

\begin{figure}[htb]
\includegraphics[width=0.5\linewidth]{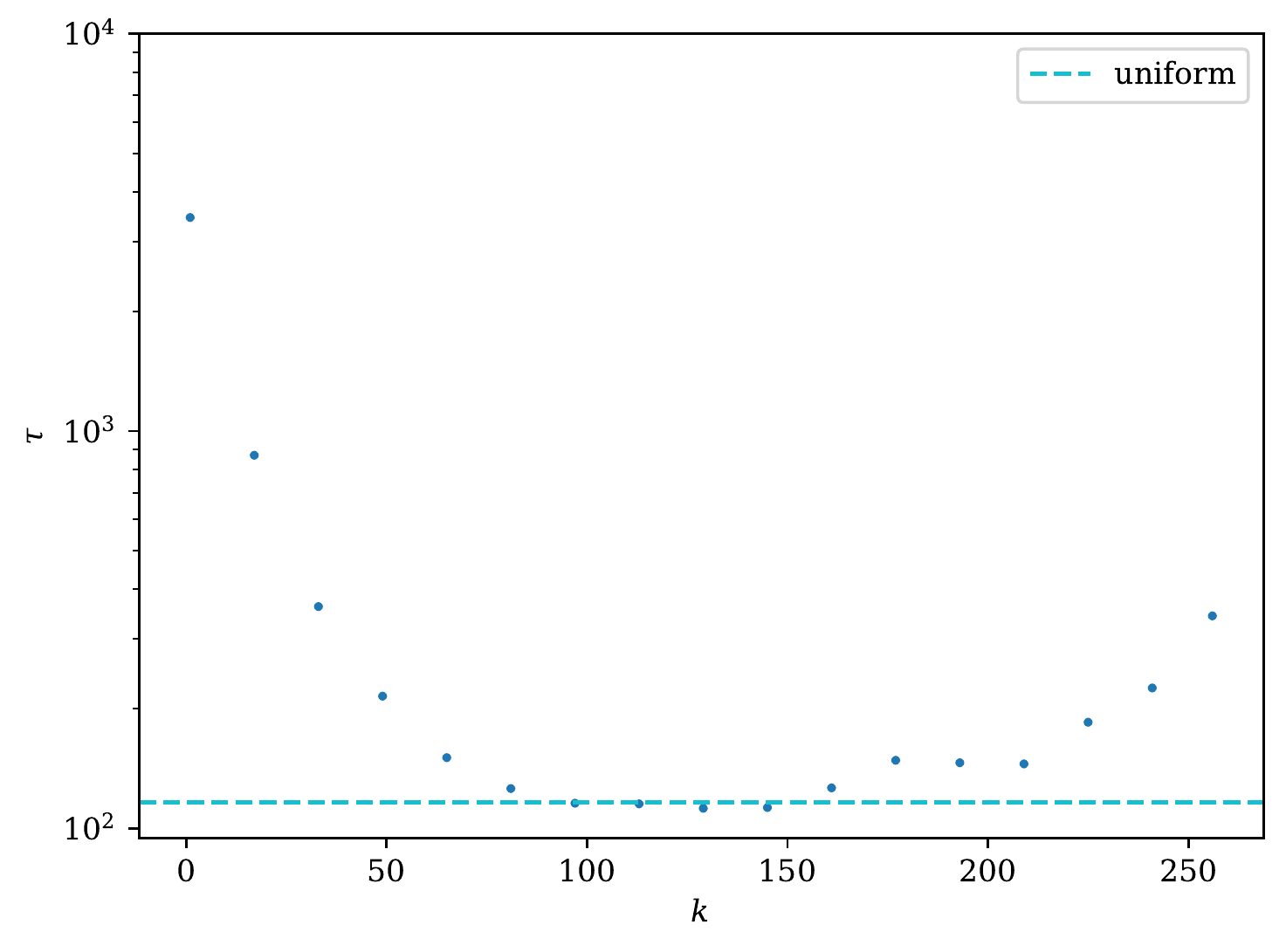}
\caption{IATs when sampling with each $k$ individually ($P_\text{cluster}(k') = \delta(k' - k)$), computed on the $16 \times 16$ Ising model. The dashed horizontal line denotes the result with the uniform distribution. Although some medium-sized $k$ give slightly lower IAT than the uniform distribution, it still performs better than most individual $k$.}
\label{fig:ising_iat_k}
\end{figure}

In NCUS, we decompose $M$ as
\begin{equation}
M = M_S M_K,
\end{equation}
where $M_S$ contains the symmetry operations, and $M_K$ represents the cluster update. Specifically, we write $M_K$ as a convex combination of different-size cluster updates
\begin{equation}
M_K = \sum_{k = 1}^V P_\text{cluster}(k) \, M^{(k)},
\end{equation}
where $\sum_k P_\text{cluster}(k) = 1$, and each of $\{M^{(k)}\}$ contains the proposals and the rejections when only the last $k$ spins are sampled. $M^{(k)}$ with a smaller $k$ usually has a higher acceptance rate. See \Fig{ising_autocorr_fun_dist_k} and \Fig{ising_iat_k} for a comparison of different choices of $P_\text{cluster}$. The sampling of $M_S$ does not need rejection, because the symmetric configurations always have the same energy as the original one. We further decompose $M_S$ as
\begin{equation}
M_S = M_{R z} M_{R d} M_{R y} M_{R x} M_{T y} M_{T x},
\end{equation}
where $M_{R x}$, $M_{R y}$, $M_{R d}$, and $M_{R z}$ contain reflections along the $x$ axis, the $y$ axis, the diagonal, and the $z$ axis respectively, and
\begin{equation}
M_{T x} = \frac{1}{L} \sum_{i = 0}^{L - 1} M_{T x; i}, \quad M_{T y} = \frac{1}{L} \sum_{i = 0}^{L - 1} M_{T y; i}
\end{equation}
represent translations by $i$ spins in the $x$ and the $y$ directions respectively. In this way, we naturally decompose the whole symmetry group into the direct product of subgroups, and avoid enumerating all the symmetry operations in sampling, while the number of symmetry operations can grow exponentially with the number of symmetry subgroups.

\newpage

\section{Comparison of cluster size distributions}

\section{Autocorrelation time}

For Markov chain-based algorithms including NCUS, NGUS and NGU, we use the \emph{integrated autocorrelation time} (IAT) $\tau$ to characterize the efficiency of the algorithm~\cite{muller1973dynamic}:
\begin{equation}
\tau = \sum_{t = 1}^{t_\text{cutoff}} r(t), \quad
r(t) = \frac{\tilde{r}(t)}{\tilde{r}(0)},
\end{equation}
with
\begin{equation}
\tilde{r}(t) = \left( \frac{1}{N - t} \sum_{i = 1}^{N - t} O(\s^{(i)}) O(\s^{(i + t)}) \right) - \bar{O}^2,
\end{equation}
where $\{\s^{(1)}, \ldots, \s^{(N)}\}$ are the samples in the Markov chain, $O$ is the observable we are interested in, and $\bar{O} = \frac{1}{N} \sum_{i = 1}^N O(\s^{(i)})$.

Because the estimation of $r(t)$ contains significant noise when $t$ becomes large, we cut off $t$ in the summation for $\tau$ when $r(t)$ crosses $0$. The resulting IAT is insensitive to the cut-off point. If we change the threshold from $0$ to $0.1$, or cut off only when $100$ consecutive $r(t)$ are lower than the threshold, the relative change of $\tau$ is $< 10\%$.

As we are drawing multiple Markov chains in parallel, we need an effective IAT to represent all of them. We first compute the variance $\Var{\bar{O}_i}$ of the observable estimator for each chain $i$:
\begin{equation}
\Var{\bar{O}_i} = \frac{2 \tau_i + 1}{n} \Var{O_i},
\end{equation}
where $n$ is the chain length, and $\Var{O_i}$ is the variance of the data in this chain. Then we compute the expectation of the observable over all chains, and propagate the variance using the fact that the chains are independent of each other:
\begin{equation}
\Var{\bar{O}} = \frac{1}{m} \sum_{i = 1}^m \Var{\bar{O}_i},
\end{equation}
where $m$ is the number of chains. Now the effective IAT $\tau_\text{eff}$ can be solved from
\begin{equation}
\Var{\bar{O}} = \frac{2 \tau_\text{eff} + 1}{m n} \Var{O},
\end{equation}
where $\Var{O}$ is the variance of the data in all chains. $\tau_\text{eff}$ is independent of the number of chains or the chain length, as long as we have enough samples to obtain a converged estimation.

We define an effective autocorrelation time for NIS by using the increased variance created by the reweighting procedure~\cite{kong1992note}
\begin{equation}
2 \tau_\text{eff} + 1 = \frac{\Var{w O}}{\Var{O}}.
\end{equation}

\section{Details of numerical experiments}

Our network has $3$ convolutional layers, each with kernel size $5$. The convolutions are masked to implement the autoregressive property, as introduced in PixelCNN~\cite{van2016pixel}. The numbers of input, hidden, and output channels are $1 \to 16 \to 16 \to 1$. SiLU activations~\cite{elfwing2018sigmoid} are applied after the first and the second convolutional layers, which are reported to produce lower loss than ReLU. Sigmoid activation is applied after the third convolutional layer to restrain the output into $(0, 1)$. To be efficient in a large number of sampling steps, we keep the network to be lightweight, while its receptive field should be able to approximately cover the whole lattice. So we use dilated convolutions~\cite{yu2016multi} to expand the receptive field, and increase the dilation rate in each convolutional layer by a step size. The receptive field radius can be calculated by
\begin{equation}
\text{Receptive field radius} = \frac{1}{2} D \left( (D - 1) d + 2 \right) \frac{s - 1}{2},
\end{equation}
where $D = 3$ is the number of convolutional layers, $s = 5$ is the convolution kernel size, and $d$ is the dilation step size. For lattice sizes $L = 8, 16, 24, 32$, the dilation step sizes are $1, 2, 3, 4$ respectively. The network has $3,761$ non-masked parameters in total, regardless of the lattice size.

During training, we use Adam optimizer~\cite{kingma2014adam} with conventional learning rate $10^{-3}$, batch size $64$, and take $2 \times 10^4$ training steps. To avoid being trapped in local minima, especially at low temperatures, in the first $10^4$ steps we linearly anneal $\beta$ from $0$ to the desired value, which is reported to produce a lower loss than exponential annealing. We do not use weight regularization or gradient clipping, because the network is shallow and there is no significant instability in training.

For sampling, we generate $10^3$ Markov chains in parallel, each containing $10^5$ samples. The chains are initialized by samples from the network. The first $10^4$ samples in each chain are discarded, to make sure only the samples after thermalization are taken into account. For each experiment of NCUS up to $L = 32$, the Gelman--Rubin diagnostic~\cite{gelman1992inference} is less than $1.1$, which confirms the chains are thermalized. The IAT is less than $4 \times 10^3$, which is shorter than the remaining chain length by orders of magnitude.

\newpage

\section{Occurrence of exponentially suppressed configurations}

\begin{figure}[htb]
\includegraphics[width=0.6\linewidth]{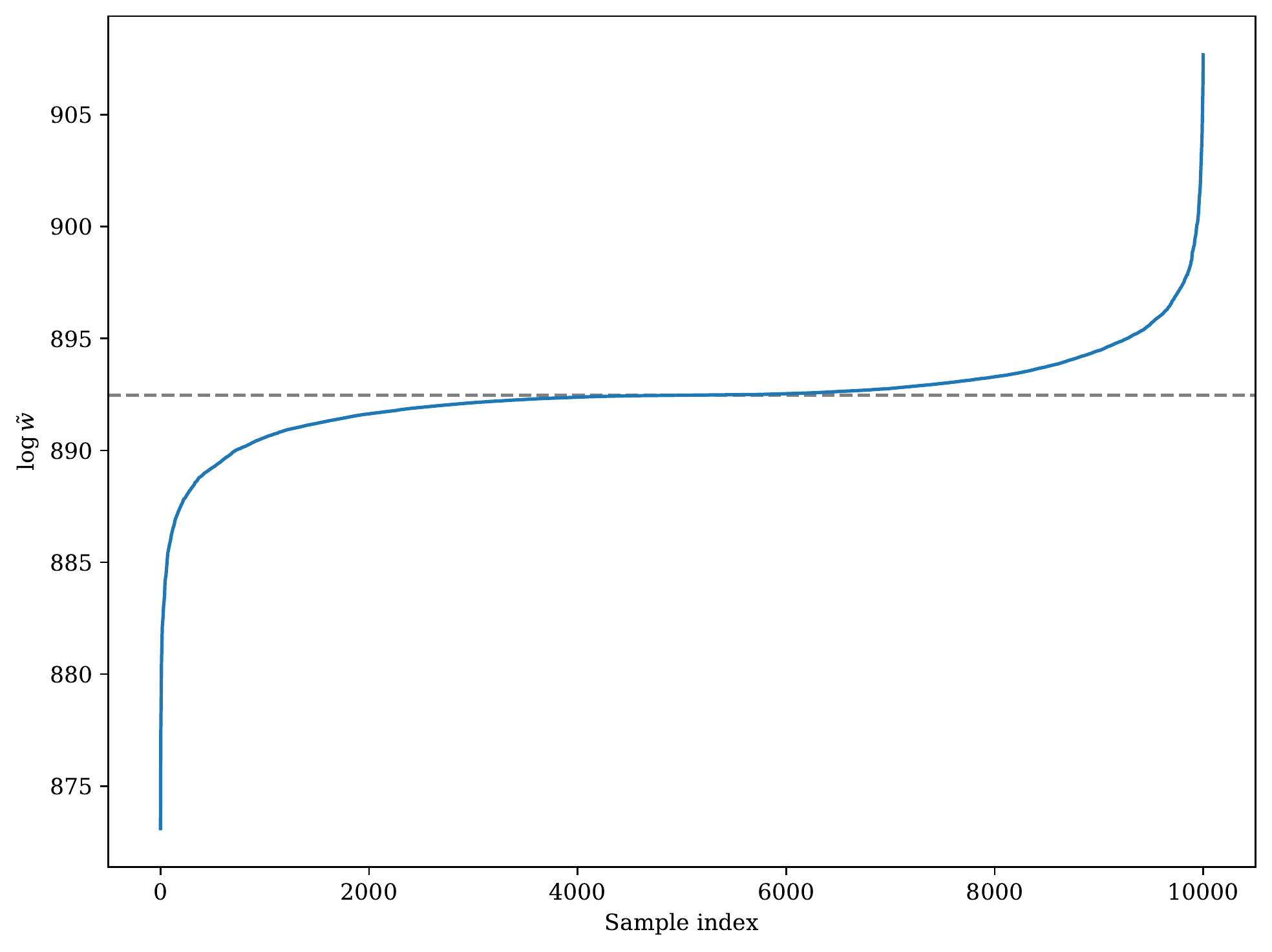}
\caption{Unnormalized importance sampling weights $\tilde{w}$ of $10^4$ random samples from a trained network on the $32 \times 32$ FPM at $\beta = 0.214$. The samples are sorted by $\tilde{w}$. The dashed horizontal line denotes the mean value of $\tilde{w}$. Only a small portion of samples have $\tilde{w}$ exponentially larger than the mean value, and they become ESC.}
\label{fig:fpm_esc}
\end{figure}

\section{System size dependence of autocorrelation time}

\begin{figure}[htb]
\centering
\includegraphics[width=0.6\linewidth]{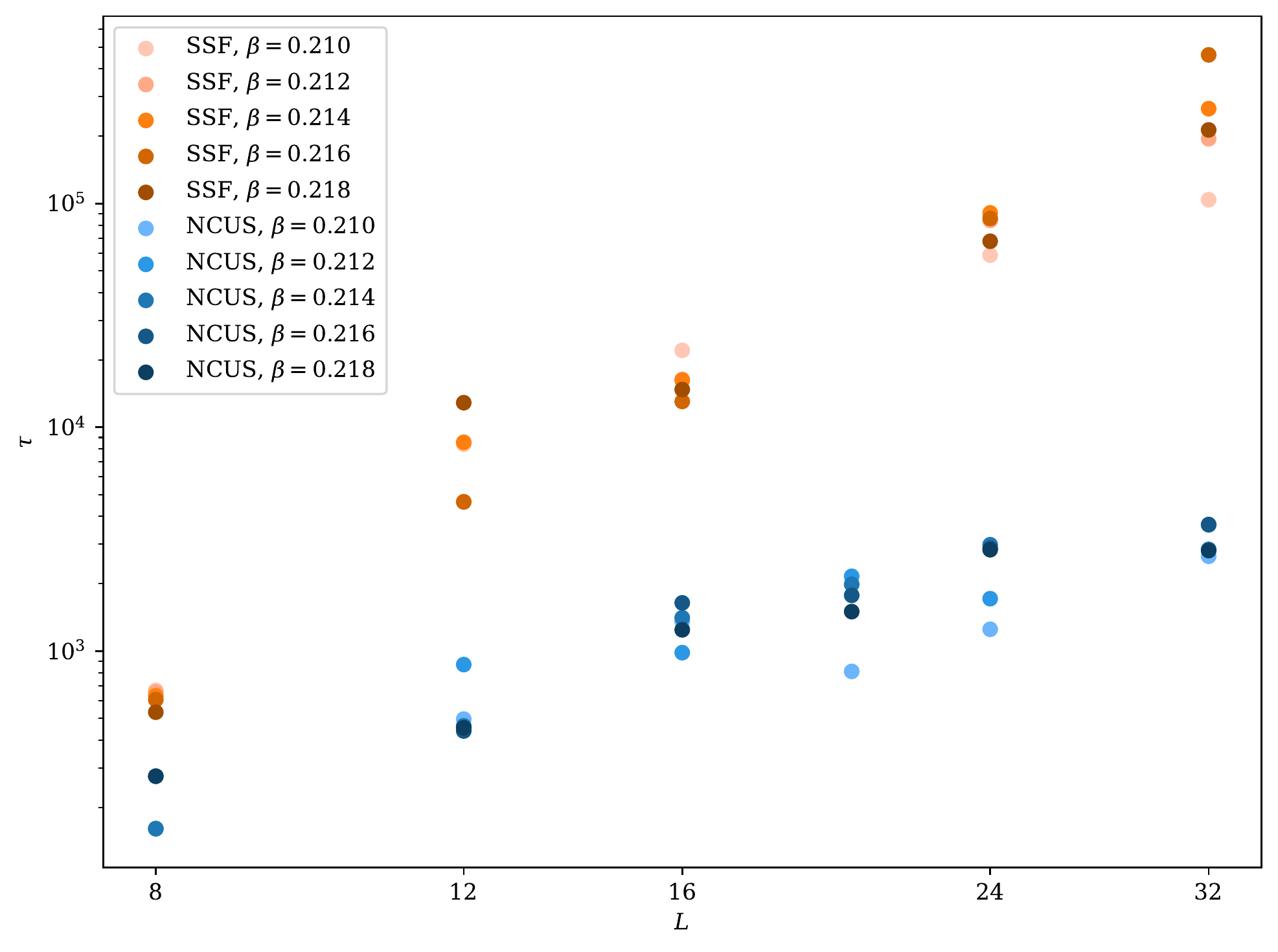}
\caption{System size dependence of the integrated autocorrelation time $\tau$ from Metropolis single spin flip method (SSF) and NCUS on FPM around the phase transition temperature. Due to technical limitations, we are currently unable to use neural networks for lattice sizes larger than $L = 32$.}
\label{fig:fpm_iat_L}
\end{figure}

\bibliography{ref}

\begin{thebibliography}{53}%
\makeatletter
\providecommand \@ifxundefined [1]{%
 \@ifx{#1\undefined}
}%
\providecommand \@ifnum [1]{%
 \ifnum #1\expandafter \@firstoftwo
 \else \expandafter \@secondoftwo
 \fi
}%
\providecommand \@ifx [1]{%
 \ifx #1\expandafter \@firstoftwo
 \else \expandafter \@secondoftwo
 \fi
}%
\providecommand \natexlab [1]{#1}%
\providecommand \enquote  [1]{``#1''}%
\providecommand \bibnamefont  [1]{#1}%
\providecommand \bibfnamefont [1]{#1}%
\providecommand \citenamefont [1]{#1}%
\providecommand \href@noop [0]{\@secondoftwo}%
\providecommand \href [0]{\begingroup \@sanitize@url \@href}%
\providecommand \@href[1]{\@@startlink{#1}\@@href}%
\providecommand \@@href[1]{\endgroup#1\@@endlink}%
\providecommand \@sanitize@url [0]{\catcode `\\12\catcode `\$12\catcode
  `\&12\catcode `\#12\catcode `\^12\catcode `\_12\catcode `\%12\relax}%
\providecommand \@@startlink[1]{}%
\providecommand \@@endlink[0]{}%
\providecommand \url  [0]{\begingroup\@sanitize@url \@url }%
\providecommand \@url [1]{\endgroup\@href {#1}{\urlprefix }}%
\providecommand \urlprefix  [0]{URL }%
\providecommand \Eprint [0]{\href }%
\providecommand \doibase [0]{https://doi.org/}%
\providecommand \selectlanguage [0]{\@gobble}%
\providecommand \bibinfo  [0]{\@secondoftwo}%
\providecommand \bibfield  [0]{\@secondoftwo}%
\providecommand \translation [1]{[#1]}%
\providecommand \BibitemOpen [0]{}%
\providecommand \bibitemStop [0]{}%
\providecommand \bibitemNoStop [0]{.\EOS\space}%
\providecommand \EOS [0]{\spacefactor3000\relax}%
\providecommand \BibitemShut  [1]{\csname bibitem#1\endcsname}%
\let\auto@bib@innerbib\@empty
\bibitem [{\citenamefont {Metropolis}\ \emph {et~al.}(1953)\citenamefont
  {Metropolis}, \citenamefont {Rosenbluth}, \citenamefont {Rosenbluth},
  \citenamefont {Teller},\ and\ \citenamefont {Teller}}]{mcmc}%
  \BibitemOpen
  \bibfield  {author} {\bibinfo {author} {\bibfnamefont {N.}~\bibnamefont
  {Metropolis}}, \bibinfo {author} {\bibfnamefont {A.~W.}\ \bibnamefont
  {Rosenbluth}}, \bibinfo {author} {\bibfnamefont {M.~N.}\ \bibnamefont
  {Rosenbluth}}, \bibinfo {author} {\bibfnamefont {A.~H.}\ \bibnamefont
  {Teller}},\ and\ \bibinfo {author} {\bibfnamefont {E.}~\bibnamefont
  {Teller}},\ }\bibfield  {title} {\bibinfo {title} {Equation of state
  calculations by fast computing machines},\ }\href@noop {} {\bibfield
  {journal} {\bibinfo  {journal} {J. Chem. Phys.}\ }\textbf {\bibinfo {volume}
  {21}},\ \bibinfo {pages} {1087} (\bibinfo {year} {1953})}\BibitemShut
  {NoStop}%
\bibitem [{\citenamefont {Binder}(1995)}]{binder1995monte}%
  \BibitemOpen
  \bibfield  {author} {\bibinfo {author} {\bibfnamefont {K.}~\bibnamefont
  {Binder}},\ }\href@noop {} {\emph {\bibinfo {title} {{M}onte {C}arlo and
  Molecular Dynamics Simulations in Polymer Science}}}\ (\bibinfo  {publisher}
  {Oxford University Press},\ \bibinfo {year} {1995})\BibitemShut {NoStop}%
\bibitem [{\citenamefont {Binder}\ and\ \citenamefont
  {Heermann}(2010)}]{binder2010monte}%
  \BibitemOpen
  \bibfield  {author} {\bibinfo {author} {\bibfnamefont {K.}~\bibnamefont
  {Binder}}\ and\ \bibinfo {author} {\bibfnamefont {D.~W.}\ \bibnamefont
  {Heermann}},\ }\href {https://doi.org/10.1007/978-3-642-03163-2} {\emph
  {\bibinfo {title} {{M}onte {C}arlo Simulation in Statistical Physics}}}\
  (\bibinfo  {publisher} {Springer},\ \bibinfo {year} {2010})\BibitemShut
  {NoStop}%
\bibitem [{\citenamefont {Krauth}(2006)}]{krauth2006statistical}%
  \BibitemOpen
  \bibfield  {author} {\bibinfo {author} {\bibfnamefont {W.}~\bibnamefont
  {Krauth}},\ }\href@noop {} {\emph {\bibinfo {title} {Statistical Mechanics:
  {A}lgorithms and Computations}}}\ (\bibinfo  {publisher} {Oxford University
  Press},\ \bibinfo {year} {2006})\BibitemShut {NoStop}%
\bibitem [{\citenamefont {Gubernatis}\ \emph {et~al.}(2016)\citenamefont
  {Gubernatis}, \citenamefont {Kawashima},\ and\ \citenamefont
  {Werner}}]{werner_book}%
  \BibitemOpen
  \bibfield  {author} {\bibinfo {author} {\bibfnamefont {J.}~\bibnamefont
  {Gubernatis}}, \bibinfo {author} {\bibfnamefont {N.}~\bibnamefont
  {Kawashima}},\ and\ \bibinfo {author} {\bibfnamefont {P.}~\bibnamefont
  {Werner}},\ }\href {https://doi.org/10.1017/9781316417041} {\emph {\bibinfo
  {title} {Quantum {M}onte {C}arlo Methods}}}\ (\bibinfo  {publisher}
  {Cambridge University Press},\ \bibinfo {year} {2016})\BibitemShut {NoStop}%
\bibitem [{\citenamefont {Becca}\ and\ \citenamefont
  {Sorella}(2017)}]{sorella_book}%
  \BibitemOpen
  \bibfield  {author} {\bibinfo {author} {\bibfnamefont {F.}~\bibnamefont
  {Becca}}\ and\ \bibinfo {author} {\bibfnamefont {S.}~\bibnamefont
  {Sorella}},\ }\href@noop {} {\emph {\bibinfo {title} {Quantum {M}onte {C}arlo
  Approaches for Correlated Systems}}}\ (\bibinfo  {publisher} {Cambridge
  University Press},\ \bibinfo {year} {2017})\BibitemShut {NoStop}%
\bibitem [{\citenamefont {Kirkpatrick}\ \emph {et~al.}(1983)\citenamefont
  {Kirkpatrick}, \citenamefont {Gelatt},\ and\ \citenamefont
  {Vecchi}}]{kirkpatrick1983optimization}%
  \BibitemOpen
  \bibfield  {author} {\bibinfo {author} {\bibfnamefont {S.}~\bibnamefont
  {Kirkpatrick}}, \bibinfo {author} {\bibfnamefont {C.~D.}\ \bibnamefont
  {Gelatt}},\ and\ \bibinfo {author} {\bibfnamefont {M.~P.}\ \bibnamefont
  {Vecchi}},\ }\bibfield  {title} {\bibinfo {title} {Optimization by simulated
  annealing},\ }\href@noop {} {\bibfield  {journal} {\bibinfo  {journal}
  {Science}\ }\textbf {\bibinfo {volume} {220}},\ \bibinfo {pages} {671}
  (\bibinfo {year} {1983})}\BibitemShut {NoStop}%
\bibitem [{\citenamefont {Rubinstein}\ and\ \citenamefont
  {Kroese}(2004)}]{rubinstein2013cross}%
  \BibitemOpen
  \bibfield  {author} {\bibinfo {author} {\bibfnamefont {R.~Y.}\ \bibnamefont
  {Rubinstein}}\ and\ \bibinfo {author} {\bibfnamefont {D.~P.}\ \bibnamefont
  {Kroese}},\ }\href@noop {} {\emph {\bibinfo {title} {The Cross-Entropy
  Method: {A} Unified Approach to Combinatorial Optimization, {M}onte-{C}arlo
  Simulation and Machine Learning}}}\ (\bibinfo  {publisher} {Springer},\
  \bibinfo {year} {2004})\BibitemShut {NoStop}%
\bibitem [{\citenamefont {Cook}(1986)}]{cook1986stochastic}%
  \BibitemOpen
  \bibfield  {author} {\bibinfo {author} {\bibfnamefont {R.~L.}\ \bibnamefont
  {Cook}},\ }\bibfield  {title} {\bibinfo {title} {Stochastic sampling in
  computer graphics},\ }\href@noop {} {\bibfield  {journal} {\bibinfo
  {journal} {ACM Trans. Graph.}\ }\textbf {\bibinfo {volume} {5}},\ \bibinfo
  {pages} {51} (\bibinfo {year} {1986})}\BibitemShut {NoStop}%
\bibitem [{\citenamefont {M{\"u}ller-Krumbhaar}\ and\ \citenamefont
  {Binder}(1973)}]{muller1973dynamic}%
  \BibitemOpen
  \bibfield  {author} {\bibinfo {author} {\bibfnamefont {H.}~\bibnamefont
  {M{\"u}ller-Krumbhaar}}\ and\ \bibinfo {author} {\bibfnamefont
  {K.}~\bibnamefont {Binder}},\ }\bibfield  {title} {\bibinfo {title} {Dynamic
  properties of the {M}onte {C}arlo method in statistical mechanics},\
  }\href@noop {} {\bibfield  {journal} {\bibinfo  {journal} {J. Stat. Phys.}\
  }\textbf {\bibinfo {volume} {8}},\ \bibinfo {pages} {1} (\bibinfo {year}
  {1973})}\BibitemShut {NoStop}%
\bibitem [{\citenamefont {Liu}(2004)}]{liu2008monte}%
  \BibitemOpen
  \bibfield  {author} {\bibinfo {author} {\bibfnamefont {J.~S.}\ \bibnamefont
  {Liu}},\ }\href@noop {} {\emph {\bibinfo {title} {{M}onte {C}arlo Strategies
  in Scientific Computing}}}\ (\bibinfo  {publisher} {Springer},\ \bibinfo
  {year} {2004})\BibitemShut {NoStop}%
\bibitem [{\citenamefont {Wang}\ and\ \citenamefont
  {Swendsen}(1990)}]{wang1990cluster}%
  \BibitemOpen
  \bibfield  {author} {\bibinfo {author} {\bibfnamefont {J.-S.}\ \bibnamefont
  {Wang}}\ and\ \bibinfo {author} {\bibfnamefont {R.~H.}\ \bibnamefont
  {Swendsen}},\ }\bibfield  {title} {\bibinfo {title} {Cluster {M}onte {C}arlo
  algorithms},\ }\href@noop {} {\bibfield  {journal} {\bibinfo  {journal}
  {Physica A}\ }\textbf {\bibinfo {volume} {167}},\ \bibinfo {pages} {565}
  (\bibinfo {year} {1990})}\BibitemShut {NoStop}%
\bibitem [{\citenamefont {Wolff}(1989)}]{wolff1989collective}%
  \BibitemOpen
  \bibfield  {author} {\bibinfo {author} {\bibfnamefont {U.}~\bibnamefont
  {Wolff}},\ }\bibfield  {title} {\bibinfo {title} {Collective {M}onte {C}arlo
  updating for spin systems},\ }\href
  {https://doi.org/10.1103/PhysRevLett.62.361} {\bibfield  {journal} {\bibinfo
  {journal} {Phys. Rev. Lett.}\ }\textbf {\bibinfo {volume} {62}},\ \bibinfo
  {pages} {361} (\bibinfo {year} {1989})}\BibitemShut {NoStop}%
\bibitem [{\citenamefont {Swendsen}\ and\ \citenamefont
  {Wang}(1986)}]{swendsen1986replica}%
  \BibitemOpen
  \bibfield  {author} {\bibinfo {author} {\bibfnamefont {R.~H.}\ \bibnamefont
  {Swendsen}}\ and\ \bibinfo {author} {\bibfnamefont {J.-S.}\ \bibnamefont
  {Wang}},\ }\bibfield  {title} {\bibinfo {title} {Replica {M}onte {C}arlo
  simulation of spin-glasses},\ }\href
  {https://doi.org/10.1103/PhysRevLett.57.2607} {\bibfield  {journal} {\bibinfo
   {journal} {Phys. Rev. Lett.}\ }\textbf {\bibinfo {volume} {57}},\ \bibinfo
  {pages} {2607} (\bibinfo {year} {1986})}\BibitemShut {NoStop}%
\bibitem [{\citenamefont {Prokof'ev}\ \emph {et~al.}(1998)\citenamefont
  {Prokof'ev}, \citenamefont {Svistunov},\ and\ \citenamefont
  {Tupitsyn}}]{prokofev1998worm}%
  \BibitemOpen
  \bibfield  {author} {\bibinfo {author} {\bibfnamefont {N.~V.}\ \bibnamefont
  {Prokof'ev}}, \bibinfo {author} {\bibfnamefont {B.~V.}\ \bibnamefont
  {Svistunov}},\ and\ \bibinfo {author} {\bibfnamefont {I.~S.}\ \bibnamefont
  {Tupitsyn}},\ }\bibfield  {title} {\bibinfo {title} {``worm'' algorithm in
  quantum {M}onte {C}arlo simulations},\ }\href@noop {} {\bibfield  {journal}
  {\bibinfo  {journal} {Phys. Lett. A}\ }\textbf {\bibinfo {volume} {238}},\
  \bibinfo {pages} {253} (\bibinfo {year} {1998})}\BibitemShut {NoStop}%
\bibitem [{\citenamefont {Bernard}\ \emph {et~al.}(2009)\citenamefont
  {Bernard}, \citenamefont {Krauth},\ and\ \citenamefont
  {Wilson}}]{ecmc_krauth}%
  \BibitemOpen
  \bibfield  {author} {\bibinfo {author} {\bibfnamefont {E.~P.}\ \bibnamefont
  {Bernard}}, \bibinfo {author} {\bibfnamefont {W.}~\bibnamefont {Krauth}},\
  and\ \bibinfo {author} {\bibfnamefont {D.~B.}\ \bibnamefont {Wilson}},\
  }\bibfield  {title} {\bibinfo {title} {Event-chain {M}onte {C}arlo algorithms
  for hard-sphere systems},\ }\href
  {https://doi.org/10.1103/PhysRevE.80.056704} {\bibfield  {journal} {\bibinfo
  {journal} {Phys. Rev. E}\ }\textbf {\bibinfo {volume} {80}},\ \bibinfo
  {pages} {056704} (\bibinfo {year} {2009})}\BibitemShut {NoStop}%
\bibitem [{\citenamefont {Carleo}\ \emph {et~al.}(2019)\citenamefont {Carleo},
  \citenamefont {Cirac}, \citenamefont {Cranmer}, \citenamefont {Daudet},
  \citenamefont {Schuld}, \citenamefont {Tishby}, \citenamefont
  {Vogt-Maranto},\ and\ \citenamefont {Zdeborov{\'a}}}]{carleo2019machine}%
  \BibitemOpen
  \bibfield  {author} {\bibinfo {author} {\bibfnamefont {G.}~\bibnamefont
  {Carleo}}, \bibinfo {author} {\bibfnamefont {I.}~\bibnamefont {Cirac}},
  \bibinfo {author} {\bibfnamefont {K.}~\bibnamefont {Cranmer}}, \bibinfo
  {author} {\bibfnamefont {L.}~\bibnamefont {Daudet}}, \bibinfo {author}
  {\bibfnamefont {M.}~\bibnamefont {Schuld}}, \bibinfo {author} {\bibfnamefont
  {N.}~\bibnamefont {Tishby}}, \bibinfo {author} {\bibfnamefont
  {L.}~\bibnamefont {Vogt-Maranto}},\ and\ \bibinfo {author} {\bibfnamefont
  {L.}~\bibnamefont {Zdeborov{\'a}}},\ }\bibfield  {title} {\bibinfo {title}
  {Machine learning and the physical sciences},\ }\href@noop {} {\bibfield
  {journal} {\bibinfo  {journal} {Rev. Mod. Phys.}\ }\textbf {\bibinfo {volume}
  {91}},\ \bibinfo {pages} {045002} (\bibinfo {year} {2019})}\BibitemShut
  {NoStop}%
\bibitem [{\citenamefont {Levy}\ \emph {et~al.}(2018)\citenamefont {Levy},
  \citenamefont {Hoffman},\ and\ \citenamefont
  {Sohl-Dickstein}}]{levy2017generalizing}%
  \BibitemOpen
  \bibfield  {author} {\bibinfo {author} {\bibfnamefont {D.}~\bibnamefont
  {Levy}}, \bibinfo {author} {\bibfnamefont {M.~D.}\ \bibnamefont {Hoffman}},\
  and\ \bibinfo {author} {\bibfnamefont {J.}~\bibnamefont {Sohl-Dickstein}},\
  }\bibfield  {title} {\bibinfo {title} {Generalizing {H}amiltonian {M}onte
  {C}arlo with neural networks},\ }in\ \href {http://arxiv.org/abs/1711.09268}
  {\emph {\bibinfo {booktitle} {International Conference on Learning
  Representations}}}\ (\bibinfo {year} {2018})\BibitemShut {NoStop}%
\bibitem [{\citenamefont {Song}\ \emph {et~al.}(2017)\citenamefont {Song},
  \citenamefont {Zhao},\ and\ \citenamefont {Ermon}}]{song2017anicemc}%
  \BibitemOpen
  \bibfield  {author} {\bibinfo {author} {\bibfnamefont {J.}~\bibnamefont
  {Song}}, \bibinfo {author} {\bibfnamefont {S.}~\bibnamefont {Zhao}},\ and\
  \bibinfo {author} {\bibfnamefont {S.}~\bibnamefont {Ermon}},\ }\bibfield
  {title} {\bibinfo {title} {{A-NICE-MC}: {A}dversarial training for {MCMC}},\
  }in\ \href {http://arxiv.org/abs/1706.07561} {\emph {\bibinfo {booktitle}
  {Advances in Neural Information Processing Systems, Long Beach, CA}}}\
  (\bibinfo {year} {2017})\ pp.\ \bibinfo {pages} {5140--5150}\BibitemShut
  {NoStop}%
\bibitem [{\citenamefont {Medvidovic}\ \emph {et~al.}(2020)\citenamefont
  {Medvidovic}, \citenamefont {Carrasquilla}, \citenamefont {Hayward},\ and\
  \citenamefont {Kulchytskyy}}]{medvidovic2020generative}%
  \BibitemOpen
  \bibfield  {author} {\bibinfo {author} {\bibfnamefont {M.}~\bibnamefont
  {Medvidovic}}, \bibinfo {author} {\bibfnamefont {J.}~\bibnamefont
  {Carrasquilla}}, \bibinfo {author} {\bibfnamefont {L.~E.}\ \bibnamefont
  {Hayward}},\ and\ \bibinfo {author} {\bibfnamefont {B.}~\bibnamefont
  {Kulchytskyy}},\ }\bibfield  {title} {\bibinfo {title} {Generative models for
  sampling of lattice field theories},\ }\href@noop {} {\  (\bibinfo {year}
  {2020})},\ \Eprint {https://arxiv.org/abs/2012.01442} {arXiv:2012.01442}
  \BibitemShut {NoStop}%
\bibitem [{\citenamefont {Liu}\ \emph {et~al.}(2017)\citenamefont {Liu},
  \citenamefont {Qi}, \citenamefont {Meng},\ and\ \citenamefont
  {Fu}}]{liu2017self}%
  \BibitemOpen
  \bibfield  {author} {\bibinfo {author} {\bibfnamefont {J.}~\bibnamefont
  {Liu}}, \bibinfo {author} {\bibfnamefont {Y.}~\bibnamefont {Qi}}, \bibinfo
  {author} {\bibfnamefont {Z.~Y.}\ \bibnamefont {Meng}},\ and\ \bibinfo
  {author} {\bibfnamefont {L.}~\bibnamefont {Fu}},\ }\bibfield  {title}
  {\bibinfo {title} {Self-learning {M}onte {C}arlo method},\ }\href
  {https://doi.org/10.1103/PhysRevB.95.041101} {\bibfield  {journal} {\bibinfo
  {journal} {Phys. Rev. B}\ }\textbf {\bibinfo {volume} {95}},\ \bibinfo
  {pages} {041101} (\bibinfo {year} {2017})}\BibitemShut {NoStop}%
\bibitem [{\citenamefont {Huang}\ and\ \citenamefont
  {Wang}(2017)}]{huang2017accelerated}%
  \BibitemOpen
  \bibfield  {author} {\bibinfo {author} {\bibfnamefont {L.}~\bibnamefont
  {Huang}}\ and\ \bibinfo {author} {\bibfnamefont {L.}~\bibnamefont {Wang}},\
  }\bibfield  {title} {\bibinfo {title} {Accelerated {M}onte {C}arlo
  simulations with restricted {B}oltzmann machines},\ }\href
  {https://doi.org/10.1103/PhysRevB.95.035105} {\bibfield  {journal} {\bibinfo
  {journal} {Phys. Rev. B}\ }\textbf {\bibinfo {volume} {95}},\ \bibinfo
  {pages} {035105} (\bibinfo {year} {2017})}\BibitemShut {NoStop}%
\bibitem [{\citenamefont {Shen}\ \emph {et~al.}(2018)\citenamefont {Shen},
  \citenamefont {Liu},\ and\ \citenamefont {Fu}}]{shen2018self}%
  \BibitemOpen
  \bibfield  {author} {\bibinfo {author} {\bibfnamefont {H.}~\bibnamefont
  {Shen}}, \bibinfo {author} {\bibfnamefont {J.}~\bibnamefont {Liu}},\ and\
  \bibinfo {author} {\bibfnamefont {L.}~\bibnamefont {Fu}},\ }\bibfield
  {title} {\bibinfo {title} {Self-learning {M}onte {C}arlo with deep neural
  networks},\ }\href {https://doi.org/10.1103/PhysRevB.97.205140} {\bibfield
  {journal} {\bibinfo  {journal} {Phys. Rev. B}\ }\textbf {\bibinfo {volume}
  {97}},\ \bibinfo {pages} {205140} (\bibinfo {year} {2018})}\BibitemShut
  {NoStop}%
\bibitem [{\citenamefont {Bonati}\ \emph {et~al.}(2019)\citenamefont {Bonati},
  \citenamefont {Zhang},\ and\ \citenamefont {Parrinello}}]{bonati2019neural}%
  \BibitemOpen
  \bibfield  {author} {\bibinfo {author} {\bibfnamefont {L.}~\bibnamefont
  {Bonati}}, \bibinfo {author} {\bibfnamefont {Y.-Y.}\ \bibnamefont {Zhang}},\
  and\ \bibinfo {author} {\bibfnamefont {M.}~\bibnamefont {Parrinello}},\
  }\bibfield  {title} {\bibinfo {title} {Neural networks-based variationally
  enhanced sampling},\ }\href {https://doi.org/10.1073/pnas.1907975116}
  {\bibfield  {journal} {\bibinfo  {journal} {Proc. Natl. Acad. Sci.}\ }\textbf
  {\bibinfo {volume} {116}},\ \bibinfo {pages} {17641} (\bibinfo {year}
  {2019})}\BibitemShut {NoStop}%
\bibitem [{\citenamefont {No{\'e}}\ \emph {et~al.}(2019)\citenamefont
  {No{\'e}}, \citenamefont {Olsson}, \citenamefont {K{\"o}hler},\ and\
  \citenamefont {Wu}}]{noe2019boltzmann}%
  \BibitemOpen
  \bibfield  {author} {\bibinfo {author} {\bibfnamefont {F.}~\bibnamefont
  {No{\'e}}}, \bibinfo {author} {\bibfnamefont {S.}~\bibnamefont {Olsson}},
  \bibinfo {author} {\bibfnamefont {J.}~\bibnamefont {K{\"o}hler}},\ and\
  \bibinfo {author} {\bibfnamefont {H.}~\bibnamefont {Wu}},\ }\bibfield
  {title} {\bibinfo {title} {Boltzmann generators: {S}ampling equilibrium
  states of many-body systems with deep learning},\ }\bibfield  {journal}
  {\bibinfo  {journal} {Science}\ }\textbf {\bibinfo {volume} {365}},\ \href
  {https://doi.org/10.1126/science.aaw1147} {10.1126/science.aaw1147} (\bibinfo
  {year} {2019})\BibitemShut {NoStop}%
\bibitem [{\citenamefont {M{\"u}ller}\ \emph {et~al.}(2019)\citenamefont
  {M{\"u}ller}, \citenamefont {McWilliams}, \citenamefont {Rousselle},
  \citenamefont {Gross},\ and\ \citenamefont {Nov{\'a}k}}]{muller2019neural}%
  \BibitemOpen
  \bibfield  {author} {\bibinfo {author} {\bibfnamefont {T.}~\bibnamefont
  {M{\"u}ller}}, \bibinfo {author} {\bibfnamefont {B.}~\bibnamefont
  {McWilliams}}, \bibinfo {author} {\bibfnamefont {F.}~\bibnamefont
  {Rousselle}}, \bibinfo {author} {\bibfnamefont {M.}~\bibnamefont {Gross}},\
  and\ \bibinfo {author} {\bibfnamefont {J.}~\bibnamefont {Nov{\'a}k}},\
  }\bibfield  {title} {\bibinfo {title} {Neural importance sampling},\
  }\href@noop {} {\bibfield  {journal} {\bibinfo  {journal} {ACM Trans.
  Graph.}\ }\textbf {\bibinfo {volume} {38}},\ \bibinfo {pages} {1} (\bibinfo
  {year} {2019})}\BibitemShut {NoStop}%
\bibitem [{\citenamefont {Wu}\ \emph {et~al.}(2019)\citenamefont {Wu},
  \citenamefont {Wang},\ and\ \citenamefont {Zhang}}]{wu2019solving}%
  \BibitemOpen
  \bibfield  {author} {\bibinfo {author} {\bibfnamefont {D.}~\bibnamefont
  {Wu}}, \bibinfo {author} {\bibfnamefont {L.}~\bibnamefont {Wang}},\ and\
  \bibinfo {author} {\bibfnamefont {P.}~\bibnamefont {Zhang}},\ }\bibfield
  {title} {\bibinfo {title} {Solving statistical mechanics using variational
  autoregressive networks},\ }\href
  {https://doi.org/10.1103/PhysRevLett.122.080602} {\bibfield  {journal}
  {\bibinfo  {journal} {Phys. Rev. Lett.}\ }\textbf {\bibinfo {volume} {122}},\
  \bibinfo {pages} {080602} (\bibinfo {year} {2019})}\BibitemShut {NoStop}%
\bibitem [{\citenamefont {Albergo}\ \emph {et~al.}(2019)\citenamefont
  {Albergo}, \citenamefont {Kanwar},\ and\ \citenamefont
  {Shanahan}}]{albergo2019flow}%
  \BibitemOpen
  \bibfield  {author} {\bibinfo {author} {\bibfnamefont {M.~S.}\ \bibnamefont
  {Albergo}}, \bibinfo {author} {\bibfnamefont {G.}~\bibnamefont {Kanwar}},\
  and\ \bibinfo {author} {\bibfnamefont {P.~E.}\ \bibnamefont {Shanahan}},\
  }\bibfield  {title} {\bibinfo {title} {Flow-based generative models for
  {M}arkov chain {M}onte {C}arlo in lattice field theory},\ }\href
  {https://doi.org/10.1103/PhysRevD.100.034515} {\bibfield  {journal} {\bibinfo
   {journal} {Phys. Rev. D}\ }\textbf {\bibinfo {volume} {100}},\ \bibinfo
  {pages} {034515} (\bibinfo {year} {2019})}\BibitemShut {NoStop}%
\bibitem [{\citenamefont {Li}\ and\ \citenamefont {Wang}(2018)}]{li2018neural}%
  \BibitemOpen
  \bibfield  {author} {\bibinfo {author} {\bibfnamefont {S.-H.}\ \bibnamefont
  {Li}}\ and\ \bibinfo {author} {\bibfnamefont {L.}~\bibnamefont {Wang}},\
  }\bibfield  {title} {\bibinfo {title} {Neural network renormalization
  group},\ }\href {https://doi.org/10.1103/PhysRevLett.121.260601} {\bibfield
  {journal} {\bibinfo  {journal} {Phys. Rev. Lett.}\ }\textbf {\bibinfo
  {volume} {121}},\ \bibinfo {pages} {260601} (\bibinfo {year}
  {2018})}\BibitemShut {NoStop}%
\bibitem [{\citenamefont {Nicoli}\ \emph {et~al.}(2020)\citenamefont {Nicoli},
  \citenamefont {Nakajima}, \citenamefont {Strodthoff}, \citenamefont {Samek},
  \citenamefont {M{\"u}ller},\ and\ \citenamefont
  {Kessel}}]{nicoli2020asymptotically}%
  \BibitemOpen
  \bibfield  {author} {\bibinfo {author} {\bibfnamefont {K.~A.}\ \bibnamefont
  {Nicoli}}, \bibinfo {author} {\bibfnamefont {S.}~\bibnamefont {Nakajima}},
  \bibinfo {author} {\bibfnamefont {N.}~\bibnamefont {Strodthoff}}, \bibinfo
  {author} {\bibfnamefont {W.}~\bibnamefont {Samek}}, \bibinfo {author}
  {\bibfnamefont {K.-R.}\ \bibnamefont {M{\"u}ller}},\ and\ \bibinfo {author}
  {\bibfnamefont {P.}~\bibnamefont {Kessel}},\ }\bibfield  {title} {\bibinfo
  {title} {Asymptotically unbiased estimation of physical observables with
  neural samplers},\ }\href {https://doi.org/10.1103/PhysRevE.101.023304}
  {\bibfield  {journal} {\bibinfo  {journal} {Phys. Rev. E}\ }\textbf {\bibinfo
  {volume} {101}},\ \bibinfo {pages} {023304} (\bibinfo {year}
  {2020})}\BibitemShut {NoStop}%
\bibitem [{\citenamefont {McNaughton}\ \emph {et~al.}(2020)\citenamefont
  {McNaughton}, \citenamefont {Milo{\v s}evi{\' c}}, \citenamefont {Perali},\
  and\ \citenamefont {Pilati}}]{mcnaughton2020boosting}%
  \BibitemOpen
  \bibfield  {author} {\bibinfo {author} {\bibfnamefont {B.}~\bibnamefont
  {McNaughton}}, \bibinfo {author} {\bibfnamefont {M.~V.}\ \bibnamefont
  {Milo{\v s}evi{\' c}}}, \bibinfo {author} {\bibfnamefont {A.}~\bibnamefont
  {Perali}},\ and\ \bibinfo {author} {\bibfnamefont {S.}~\bibnamefont
  {Pilati}},\ }\bibfield  {title} {\bibinfo {title} {Boosting {M}onte {C}arlo
  simulations of spin glasses using autoregressive neural networks},\ }\href
  {https://doi.org/10.1103/PhysRevE.101.053312} {\bibfield  {journal} {\bibinfo
   {journal} {Phys. Rev. E}\ }\textbf {\bibinfo {volume} {101}},\ \bibinfo
  {pages} {053312} (\bibinfo {year} {2020})}\BibitemShut {NoStop}%
\bibitem [{\citenamefont {Uria}\ \emph {et~al.}(2016)\citenamefont {Uria},
  \citenamefont {C{\^o}t{\'e}}, \citenamefont {Gregor}, \citenamefont
  {Murray},\ and\ \citenamefont {Larochelle}}]{uria2016neural}%
  \BibitemOpen
  \bibfield  {author} {\bibinfo {author} {\bibfnamefont {B.}~\bibnamefont
  {Uria}}, \bibinfo {author} {\bibfnamefont {M.-A.}\ \bibnamefont
  {C{\^o}t{\'e}}}, \bibinfo {author} {\bibfnamefont {K.}~\bibnamefont
  {Gregor}}, \bibinfo {author} {\bibfnamefont {I.}~\bibnamefont {Murray}},\
  and\ \bibinfo {author} {\bibfnamefont {H.}~\bibnamefont {Larochelle}},\
  }\bibfield  {title} {\bibinfo {title} {Neural autoregressive distribution
  estimation},\ }\href {https://jmlr.org/papers/v17/16-272.html} {\bibfield
  {journal} {\bibinfo  {journal} {J. Mach. Learn. Res.}\ }\textbf {\bibinfo
  {volume} {17}},\ \bibinfo {pages} {7184} (\bibinfo {year}
  {2016})}\BibitemShut {NoStop}%
\bibitem [{\citenamefont {Larochelle}\ and\ \citenamefont
  {Murray}(2011)}]{pmlr-v15-larochelle11a}%
  \BibitemOpen
  \bibfield  {author} {\bibinfo {author} {\bibfnamefont {H.}~\bibnamefont
  {Larochelle}}\ and\ \bibinfo {author} {\bibfnamefont {I.}~\bibnamefont
  {Murray}},\ }\bibfield  {title} {\bibinfo {title} {The neural autoregressive
  distribution estimator},\ }in\ \href
  {http://proceedings.mlr.press/v15/larochelle11a.html} {\emph {\bibinfo
  {booktitle} {International Conference on Artificial Intelligence and
  Statistics}}},\ Vol.~\bibinfo {volume} {15}\ (\bibinfo {year} {2011})\ pp.\
  \bibinfo {pages} {29--37}\BibitemShut {NoStop}%
\bibitem [{\citenamefont {Kullback}\ and\ \citenamefont
  {Leibler}(1951)}]{kullback1951information}%
  \BibitemOpen
  \bibfield  {author} {\bibinfo {author} {\bibfnamefont {S.}~\bibnamefont
  {Kullback}}\ and\ \bibinfo {author} {\bibfnamefont {R.~A.}\ \bibnamefont
  {Leibler}},\ }\bibfield  {title} {\bibinfo {title} {On information and
  sufficiency},\ }\href@noop {} {\bibfield  {journal} {\bibinfo  {journal}
  {Ann. Math. Stat.}\ }\textbf {\bibinfo {volume} {22}},\ \bibinfo {pages} {79}
  (\bibinfo {year} {1951})}\BibitemShut {NoStop}%
\bibitem [{\citenamefont {Goodfellow}\ \emph {et~al.}(2016)\citenamefont
  {Goodfellow}, \citenamefont {Bengio},\ and\ \citenamefont
  {Courville}}]{goodfellow2016deep}%
  \BibitemOpen
  \bibfield  {author} {\bibinfo {author} {\bibfnamefont {I.}~\bibnamefont
  {Goodfellow}}, \bibinfo {author} {\bibfnamefont {Y.}~\bibnamefont {Bengio}},\
  and\ \bibinfo {author} {\bibfnamefont {A.}~\bibnamefont {Courville}},\
  }\href@noop {} {\emph {\bibinfo {title} {Deep Learning}}}\ (\bibinfo
  {publisher} {MIT Press},\ \bibinfo {year} {2016})\ Chap.\ \bibinfo {chapter}
  {3.1.3}\BibitemShut {NoStop}%
\bibitem [{sm()}]{sm}%
  \BibitemOpen
  \href@noop {} {}\bibinfo {note} {See Supplemental Material at [URL will be
  inserted by publisher] for a discussion about the decomposition of transition
  matrix, a comparison of cluster size distributions $P_\text{cluster}$, the
  definitions of autocorrelation times, details of numerical experiments, a
  plot of the occurrence of exponentially suppressed configurations (ESC), and
  the system size dependence of autocorrelation time.}\BibitemShut {Stop}%
\bibitem [{\citenamefont {Onsager}(1944)}]{onsager1944crystal}%
  \BibitemOpen
  \bibfield  {author} {\bibinfo {author} {\bibfnamefont {L.}~\bibnamefont
  {Onsager}},\ }\bibfield  {title} {\bibinfo {title} {Crystal statistics. {I}.
  {A} two-dimensional model with an order-disorder transition},\ }\href
  {https://doi.org/10.1103/PhysRev.65.117} {\bibfield  {journal} {\bibinfo
  {journal} {Phys. Rev.}\ }\textbf {\bibinfo {volume} {65}},\ \bibinfo {pages}
  {117} (\bibinfo {year} {1944})}\BibitemShut {NoStop}%
\bibitem [{\citenamefont {van~den Oord}\ \emph {et~al.}(2016)\citenamefont
  {van~den Oord}, \citenamefont {Kalchbrenner},\ and\ \citenamefont
  {Kavukcuoglu}}]{van2016pixel}%
  \BibitemOpen
  \bibfield  {author} {\bibinfo {author} {\bibfnamefont {A.}~\bibnamefont
  {van~den Oord}}, \bibinfo {author} {\bibfnamefont {N.}~\bibnamefont
  {Kalchbrenner}},\ and\ \bibinfo {author} {\bibfnamefont {K.}~\bibnamefont
  {Kavukcuoglu}},\ }\bibfield  {title} {\bibinfo {title} {Pixel recurrent
  neural networks},\ }in\ \href@noop {} {\emph {\bibinfo {booktitle}
  {International Conference on Machine Learning}}}\ (\bibinfo {year}
  {2016})\BibitemShut {NoStop}%
\bibitem [{\citenamefont {Yu}\ and\ \citenamefont
  {Koltun}(2016)}]{yu2016multi}%
  \BibitemOpen
  \bibfield  {author} {\bibinfo {author} {\bibfnamefont {F.}~\bibnamefont
  {Yu}}\ and\ \bibinfo {author} {\bibfnamefont {V.}~\bibnamefont {Koltun}},\
  }\bibfield  {title} {\bibinfo {title} {Multi-scale context aggregation by
  dilated convolutions},\ }in\ \href {http://arxiv.org/abs/1511.07122} {\emph
  {\bibinfo {booktitle} {International Conference on Learning
  Representations}}}\ (\bibinfo {year} {2016})\BibitemShut {NoStop}%
\bibitem [{\citenamefont {Vollmayr}\ \emph {et~al.}(1993)\citenamefont
  {Vollmayr}, \citenamefont {Reger}, \citenamefont {Scheucher},\ and\
  \citenamefont {Binder}}]{vollmayr1993finite}%
  \BibitemOpen
  \bibfield  {author} {\bibinfo {author} {\bibfnamefont {K.}~\bibnamefont
  {Vollmayr}}, \bibinfo {author} {\bibfnamefont {J.~D.}\ \bibnamefont {Reger}},
  \bibinfo {author} {\bibfnamefont {M.}~\bibnamefont {Scheucher}},\ and\
  \bibinfo {author} {\bibfnamefont {K.}~\bibnamefont {Binder}},\ }\bibfield
  {title} {\bibinfo {title} {Finite size effects at thermally-driven first
  order phase transitions: {A} phenomenological theory of the order parameter
  distribution},\ }\href@noop {} {\bibfield  {journal} {\bibinfo  {journal} {Z.
  Phys., B Condens. Matter}\ }\textbf {\bibinfo {volume} {91}},\ \bibinfo
  {pages} {113} (\bibinfo {year} {1993})}\BibitemShut {NoStop}%
\bibitem [{\citenamefont {Arisue}\ and\ \citenamefont
  {Tabata}(2001)}]{arisue2001first}%
  \BibitemOpen
  \bibfield  {author} {\bibinfo {author} {\bibfnamefont {H.}~\bibnamefont
  {Arisue}}\ and\ \bibinfo {author} {\bibfnamefont {K.}~\bibnamefont
  {Tabata}},\ }\bibfield  {title} {\bibinfo {title} {First order phase
  transition of the $q$-state {P}otts model in two dimensions},\ }in\
  \href@noop {} {\emph {\bibinfo {booktitle} {Non-Perturbative Methods and
  Lattice QCD}}}\ (\bibinfo  {publisher} {World Scientific},\ \bibinfo {year}
  {2001})\ pp.\ \bibinfo {pages} {233--241}\BibitemShut {NoStop}%
\bibitem [{\citenamefont {Gorbenko}\ \emph
  {et~al.}(2018{\natexlab{a}})\citenamefont {Gorbenko}, \citenamefont
  {Rychkov},\ and\ \citenamefont {Zan}}]{gorbenko2018walking}%
  \BibitemOpen
  \bibfield  {author} {\bibinfo {author} {\bibfnamefont {V.}~\bibnamefont
  {Gorbenko}}, \bibinfo {author} {\bibfnamefont {S.}~\bibnamefont {Rychkov}},\
  and\ \bibinfo {author} {\bibfnamefont {B.}~\bibnamefont {Zan}},\ }\bibfield
  {title} {\bibinfo {title} {Walking, weak first-order transitions, and complex
  {CFT}s},\ }\href@noop {} {\bibfield  {journal} {\bibinfo  {journal} {J. High
  Energy Phys.}\ }\textbf {\bibinfo {volume} {2018}}\bibinfo  {number} {
  (10)},\ \bibinfo {pages} {1}}\BibitemShut {NoStop}%
\bibitem [{\citenamefont {Gorbenko}\ \emph
  {et~al.}(2018{\natexlab{b}})\citenamefont {Gorbenko}, \citenamefont
  {Rychkov},\ and\ \citenamefont {Zan}}]{gorbenko2018walking2}%
  \BibitemOpen
\bibfield  {number} {  }\bibfield  {author} {\bibinfo {author} {\bibfnamefont
  {V.}~\bibnamefont {Gorbenko}}, \bibinfo {author} {\bibfnamefont
  {S.}~\bibnamefont {Rychkov}},\ and\ \bibinfo {author} {\bibfnamefont
  {B.}~\bibnamefont {Zan}},\ }\bibfield  {title} {\bibinfo {title} {Walking,
  weak first-order transitions, and complex {CFT}s {II}. {T}wo-dimensional
  {P}otts model at ${Q} > 4$},\ }\href
  {https://doi.org/10.21468/SciPostPhys.5.5.050} {\bibfield  {journal}
  {\bibinfo  {journal} {SciPost Phys.}\ }\textbf {\bibinfo {volume} {5}},\
  \bibinfo {pages} {050} (\bibinfo {year} {2018}{\natexlab{b}})}\BibitemShut
  {NoStop}%
\bibitem [{\citenamefont {Gheissari}\ and\ \citenamefont
  {Lubetzky}(2016)}]{gheissari2016mixing}%
  \BibitemOpen
  \bibfield  {author} {\bibinfo {author} {\bibfnamefont {R.}~\bibnamefont
  {Gheissari}}\ and\ \bibinfo {author} {\bibfnamefont {E.}~\bibnamefont
  {Lubetzky}},\ }\bibfield  {title} {\bibinfo {title} {Mixing times of critical
  {2D} {P}otts models},\ }\href@noop {} {\  (\bibinfo {year} {2016})},\ \Eprint
  {https://arxiv.org/abs/1607.02182} {arXiv:1607.02182} \BibitemShut {NoStop}%
\bibitem [{\citenamefont {Krzakala}\ and\ \citenamefont
  {Zdeborov{\'a}}(2009)}]{krzakala2009hiding}%
  \BibitemOpen
  \bibfield  {author} {\bibinfo {author} {\bibfnamefont {F.}~\bibnamefont
  {Krzakala}}\ and\ \bibinfo {author} {\bibfnamefont {L.}~\bibnamefont
  {Zdeborov{\'a}}},\ }\bibfield  {title} {\bibinfo {title} {Hiding quiet
  solutions in random constraint satisfaction problems},\ }\href
  {https://doi.org/10.1103/PhysRevLett.102.238701} {\bibfield  {journal}
  {\bibinfo  {journal} {Phys. Rev. Lett.}\ }\textbf {\bibinfo {volume} {102}},\
  \bibinfo {pages} {238701} (\bibinfo {year} {2009})}\BibitemShut {NoStop}%
\bibitem [{\citenamefont {Krzakala}\ \emph {et~al.}(2012)\citenamefont
  {Krzakala}, \citenamefont {M{\'e}zard}, \citenamefont {Sausset},
  \citenamefont {Sun},\ and\ \citenamefont
  {Zdeborov{\'a}}}]{krzakala2012statistical}%
  \BibitemOpen
  \bibfield  {author} {\bibinfo {author} {\bibfnamefont {F.}~\bibnamefont
  {Krzakala}}, \bibinfo {author} {\bibfnamefont {M.}~\bibnamefont
  {M{\'e}zard}}, \bibinfo {author} {\bibfnamefont {F.}~\bibnamefont {Sausset}},
  \bibinfo {author} {\bibfnamefont {Y.~F.}\ \bibnamefont {Sun}},\ and\ \bibinfo
  {author} {\bibfnamefont {L.}~\bibnamefont {Zdeborov{\'a}}},\ }\bibfield
  {title} {\bibinfo {title} {Statistical-physics-based reconstruction in
  compressed sensing},\ }\href {https://doi.org/10.1103/PhysRevX.2.021005}
  {\bibfield  {journal} {\bibinfo  {journal} {Phys. Rev. X}\ }\textbf {\bibinfo
  {volume} {2}},\ \bibinfo {pages} {021005} (\bibinfo {year}
  {2012})}\BibitemShut {NoStop}%
\bibitem [{\citenamefont {Banks}\ \emph {et~al.}(2019)\citenamefont {Banks},
  \citenamefont {Kleinberg},\ and\ \citenamefont {Moore}}]{banks2017lovasz}%
  \BibitemOpen
  \bibfield  {author} {\bibinfo {author} {\bibfnamefont {J.}~\bibnamefont
  {Banks}}, \bibinfo {author} {\bibfnamefont {R.}~\bibnamefont {Kleinberg}},\
  and\ \bibinfo {author} {\bibfnamefont {C.}~\bibnamefont {Moore}},\ }\bibfield
   {title} {\bibinfo {title} {The {L}ov{\'a}sz theta function for random
  regular graphs and community detection in the hard regime},\ }\bibfield
  {journal} {\bibinfo  {journal} {SIAM J. Comput.}\ }\textbf {\bibinfo {volume}
  {48}},\ \href {https://doi.org/10.1137/18M1180396} {10.1137/18M1180396}
  (\bibinfo {year} {2019})\BibitemShut {NoStop}%
\bibitem [{\citenamefont {Blankenbecler}\ \emph {et~al.}(1981)\citenamefont
  {Blankenbecler}, \citenamefont {Scalapino},\ and\ \citenamefont
  {Sugar}}]{blankenbecler1981monte}%
  \BibitemOpen
  \bibfield  {author} {\bibinfo {author} {\bibfnamefont {R.}~\bibnamefont
  {Blankenbecler}}, \bibinfo {author} {\bibfnamefont {D.~J.}\ \bibnamefont
  {Scalapino}},\ and\ \bibinfo {author} {\bibfnamefont {R.~L.}\ \bibnamefont
  {Sugar}},\ }\bibfield  {title} {\bibinfo {title} {{M}onte {C}arlo
  calculations of coupled boson-fermion systems. {I}},\ }\href
  {https://doi.org/10.1103/PhysRevD.24.2278} {\bibfield  {journal} {\bibinfo
  {journal} {Phys. Rev. D}\ }\textbf {\bibinfo {volume} {24}},\ \bibinfo
  {pages} {2278} (\bibinfo {year} {1981})}\BibitemShut {NoStop}%
\bibitem [{\citenamefont {Hackett}\ \emph {et~al.}(2021)\citenamefont
  {Hackett}, \citenamefont {Hsieh}, \citenamefont {Albergo}, \citenamefont
  {Boyda}, \citenamefont {Chen}, \citenamefont {Chen}, \citenamefont {Cranmer},
  \citenamefont {Kanwar},\ and\ \citenamefont {Shanahan}}]{hackett2021flow}%
  \BibitemOpen
  \bibfield  {author} {\bibinfo {author} {\bibfnamefont {D.~C.}\ \bibnamefont
  {Hackett}}, \bibinfo {author} {\bibfnamefont {C.-C.}\ \bibnamefont {Hsieh}},
  \bibinfo {author} {\bibfnamefont {M.~S.}\ \bibnamefont {Albergo}}, \bibinfo
  {author} {\bibfnamefont {D.}~\bibnamefont {Boyda}}, \bibinfo {author}
  {\bibfnamefont {J.-W.}\ \bibnamefont {Chen}}, \bibinfo {author}
  {\bibfnamefont {K.-F.}\ \bibnamefont {Chen}}, \bibinfo {author}
  {\bibfnamefont {K.}~\bibnamefont {Cranmer}}, \bibinfo {author} {\bibfnamefont
  {G.}~\bibnamefont {Kanwar}},\ and\ \bibinfo {author} {\bibfnamefont {P.~E.}\
  \bibnamefont {Shanahan}},\ }\bibfield  {title} {\bibinfo {title} {Flow-based
  sampling for multimodal distributions in lattice field theory},\ }\href@noop
  {} {\  (\bibinfo {year} {2021})},\ \Eprint {https://arxiv.org/abs/2107.00734}
  {arxiv:2107.00734} \BibitemShut {NoStop}%
\bibitem [{\citenamefont {Kong}(1992)}]{kong1992note}%
  \BibitemOpen
  \bibfield  {author} {\bibinfo {author} {\bibfnamefont {A.}~\bibnamefont
  {Kong}},\ }\bibfield  {title} {\bibinfo {title} {A note on importance
  sampling using standardized weights},\ }in\ \href@noop {} {\emph {\bibinfo
  {booktitle} {University of Chicago, Dept. of Statistics, Tech. Rep.}}},\
  Vol.\ \bibinfo {volume} {348}\ (\bibinfo {year} {1992})\BibitemShut {NoStop}%
\bibitem [{\citenamefont {Elfwing}\ \emph {et~al.}(2018)\citenamefont
  {Elfwing}, \citenamefont {Uchibe},\ and\ \citenamefont
  {Doya}}]{elfwing2018sigmoid}%
  \BibitemOpen
  \bibfield  {author} {\bibinfo {author} {\bibfnamefont {S.}~\bibnamefont
  {Elfwing}}, \bibinfo {author} {\bibfnamefont {E.}~\bibnamefont {Uchibe}},\
  and\ \bibinfo {author} {\bibfnamefont {K.}~\bibnamefont {Doya}},\ }\bibfield
  {title} {\bibinfo {title} {Sigmoid-weighted linear units for neural network
  function approximation in reinforcement learning},\ }\href@noop {} {\bibfield
   {journal} {\bibinfo  {journal} {Neural Netw.}\ }\textbf {\bibinfo {volume}
  {107}},\ \bibinfo {pages} {3} (\bibinfo {year} {2018})}\BibitemShut {NoStop}%
\bibitem [{\citenamefont {Kingma}\ and\ \citenamefont
  {Ba}(2015)}]{kingma2014adam}%
  \BibitemOpen
  \bibfield  {author} {\bibinfo {author} {\bibfnamefont {D.~P.}\ \bibnamefont
  {Kingma}}\ and\ \bibinfo {author} {\bibfnamefont {J.}~\bibnamefont {Ba}},\
  }\bibfield  {title} {\bibinfo {title} {Adam: {A} method for stochastic
  optimization},\ }in\ \href@noop {} {\emph {\bibinfo {booktitle}
  {International Conference on Machine Learning}}}\ (\bibinfo {year}
  {2015})\BibitemShut {NoStop}%
\bibitem [{\citenamefont {Gelman}\ and\ \citenamefont
  {Rubin}(1992)}]{gelman1992inference}%
  \BibitemOpen
  \bibfield  {author} {\bibinfo {author} {\bibfnamefont {A.}~\bibnamefont
  {Gelman}}\ and\ \bibinfo {author} {\bibfnamefont {D.~B.}\ \bibnamefont
  {Rubin}},\ }\bibfield  {title} {\bibinfo {title} {Inference from iterative
  simulation using multiple sequences},\ }\href@noop {} {\bibfield  {journal}
  {\bibinfo  {journal} {Stat. Sci.}\ }\textbf {\bibinfo {volume} {7}},\
  \bibinfo {pages} {457} (\bibinfo {year} {1992})}\BibitemShut {NoStop}%
\end{thebibliography}%

\end{document}